\documentclass[onecolumn,tighten,times]{aastex62}

\usepackage{graphicx}
\usepackage{natbib}

\catcode`\@=11
\newcommand{\gapprox}{\mathrel{\mathpalette\@versim>}}
\newcommand{\lapprox}{\mathrel{\mathpalette\@versim<}}
\newcommand{\propapprox}{\mathrel{\mathpalette\@versim\propto}}
\newcommand{\@versim}[2]
  {\lower3.1truept\vbox{\baselineskip0pt\lineskip0.5truept
\ialign{$\m@th#1\hfil##\hfil$\crcr#2\crcr\sim\crcr}}}
\catcode`\@=12

\shorttitle{NONTHERMAL EFFICIENCIES IN KEPLER'S SNR}

\begin{document}

\title{Efficiencies of Magnetic-Field Amplification and Electron
  Acceleration in Young Supernova Remnants:  Global Averages and Kepler's Supernova Remnant}

\email{reynolds@ncsu.edu}

\author{Stephen P.~Reynolds}
\affiliation{Department of Physics,
    North Carolina State University, \\ Raleigh, NC 27695-8202}

\author{Brian J.~Williams}
    \affiliation{NASA/GSFC, Code 660,
    Greenbelt, MD 20771}

\author{Kazimierz J.~Borkowski}
    \affiliation{Department of Physics,
    North Carolina State University, Raleigh, NC 27695-8202}

\author{Knox S.~Long} \affiliation{Space Telescope Science Institute,
  3700 San Martin Drive, Baltimore, MD, 21218}


\begin{abstract}

Particle acceleration to suprathermal energies in strong astrophysical
shock waves is a widespread phenomenon, generally explained by
diffusive shock acceleration.  Such shocks can also amplify upstream
magnetic field considerably beyond simple compression. The complex
plasma physics processes involved are often parameterized by assuming
that shocks put some fraction $\epsilon_e$ of their energy into fast
particles, and another fraction $\epsilon_B$ into magnetic
field. Modelers of shocks in supernovae, supernova remnants, and
gamma-ray bursters, among other locations, often assume typical
values for these fractions, presumed to remain constant in time.
However, it is rare that enough properties of a source are
independently constrained that values of the epsilons can be inferred
directly. Supernova remnants (SNRs) can provide such circumstances.
Here we summarize results from global fits to spatially integrated
emission in six young SNRs, finding $10^{-4} \lapprox \epsilon_e
\lapprox 0.05$ and $0.001 \lapprox \epsilon_B \lapprox 0.1$.  These
large variations might be put down to the differing ages and
environments of these SNRs, so we conduct a detailed analysis of a
single remnant, that of Kepler's supernova. Both epsilons can be
determined at seven different locations around the shock, and we find
even larger ranges for both epsilons, as well as for their ratio (thus
independent of the shock energy itself).  We conclude that unknown
factors have a large influence on the efficiency of both processes.
Shock obliquity, upstream neutral fraction, or other possibilities
need to be explored, while calculations assuming fixed values of the
epsilons should be regarded as provisional.

\end{abstract}

\section{Introduction}

Strong shocks in supernovae, supernova remnants (SNRs), and gamma-ray
bursters (GRB sources) are widely observed to produce nonthermal
particle distributions and to amplify ambient magnetic fields
\cite[e.g.,][]{weiler02,reynolds08, vanparadijs00}. The total energy in
these nonthermal components is generally small compared to bulk
thermal and kinetic energies; to produce the Galactic cosmic rays,
roughly 10\% of supernova energy is adequate. The processes
responsible for particle acceleration and magnetic-field amplification
are fairly well understood in broad terms \citep[see any of various reviews,
  such as][]{blandford87,malkov01,ressler14}
but details remain frustratingly elusive.  Even as
basic an issue as the fraction of shock energy eventually
winding up in particles and magnetic field is difficult to predict
from first principles, and may evolve with changing conditions.

Observational constraints on these quantities are also surprisingly
hard to come by.  An early attempt at using the spatial structure of
radio images of young supernova remnants (SNRs) to discriminate among
models for magnetic-field evolution \citep{reynolds81} compared
theoretical profiles of SNR radio emission assuming uniform post-shock
magnetic-field strength (shock expanding into constant magnetic field,
with only compression increasing the post-shock value), with a model
with magnetic energy density $u_B$ amplified to a value proportional
to post-shock pressure $P_2$.  (The post-shock pressure in a strong
shock from the Rankine-Hugoniot jump conditions is $P_2 = [2/(\gamma +
  1)]\rho_0 v_s^2$ or $3\rho_0 v_s^2/4$ for adiabatic index $\gamma =
5/3$, where $\rho_0$ is the upstream gas density and $v_s$ the shock
speed. $P_2$ is also proportional to the total thermal energy density
$e_2$: $e_2 = P_2/(\gamma - 1)$, so that $\rho_0 v_s^2 = [(\gamma +
  1)(\gamma -1)/2] e_2 = (8/9)e_2$ for $\gamma = 5/3$.  The convention
has been to absorb the factor $8/9$ into the definitions of the
$\epsilon$ efficiency factors, and we shall follow that convention
here: $\epsilon_e \equiv u_e/\rho_0 v_s^2$ and $\epsilon_B \equiv
(B^2/8\pi)/\rho_0 v_s^2$.)

\cite{reynolds81} found substantial differences between predictions of their two
models for magnetic field: amplified-field models predicted that the
radio profile should rise from the shock inward all the way to the
contact discontinuity between shocked ejecta and shocked ambient
material.  However, various confounding effects made this prediction
difficult to test.  A similar test \citep{chomiuk09} used collective
luminosity functions of extragalactic SNRs to try to discriminate
between these models, finding that (assuming a constant cosmic-ray
energy density) the swept-up model was less favored by the data.
However, the assumption of a non-evolving relativistic-electron
density is not supported by more recent studies of diffusive shock
acceleration (DSA).  See, for instance, \cite{riquelme11}.

The dependence of the synchrotron luminosity on both the energy
density of relativistic electrons and of magnetic field
plagues attempts to use synchrotron emission as a diagnostic in
contexts as disparate as active galactic nuclei, gamma-ray burst
sources, the interstellar medium of normal galaxies, supernovae, nova
and supernova remnants, pulsar-wind nebulae, and elsewhere.  By now
the assumption of equipartition of energy between particles and field
has been largely abandoned.  Originally introduced to obtain a lower
limit to the energy required to explain extended emission in radio
galaxies \citep[in particular M87;][]{burbidge56}, the assumption took
on a life of its own, supported by no more than a vague sense that
Nature ought to put equal amounts of energy in each available form.
The difficulty of deciding how much energy might reside in
relativistic protons has meant that apart from providing absolute
lower limits for energy arguments, in which case one ignores energy in
protons, the assumption has substantial model dependence, and its
utility and predictive power are limited.

The assumption of constant fractions of shock energy going into
relativistic electrons ($\epsilon_e$) and magnetic field
($\epsilon_B$) is implicit in \cite{reynolds81}, and made explicitly in
\cite{chevalier84}.  It appears to have been the standard assumption
in early GRB modeling, which seems to be where the $\epsilon$ notation
originated \citep[e.g.,][]{sari98}.  The notation, and the assumption
that these fractions remain constant as the shock wave evolves, have
become widespread in modeling radio emission from supernovae and SNRs
as well as GRBs \citep[e.g.,][]{lundqvist20}.  Typical assumed values
for these fractions are in the range 0.01 to 0.1 for GRBs
\citep[e.g.,][]{nava14}.  However, \cite{panaitescu01} fit a
multiparameter model for GRB afterglows to 8 events, finding values
ranging over an order of magnitude for $\epsilon_e$ ($10^{-2} -
10^{-1}$) and three orders of magnitude for $\epsilon_B$ ($10^{-4} -
10^{-1}$).  In general, both the values and the assumption of
constancy lack the observational support one would like in such
fundamental quantities.  Additionally, the relativistic shocks
inferred in GRB afterglows may have qualitatively different effects in
both magnetic-field amplification and electron acceleration than would
be found in nonrelativistic shocks in SNRs.  Numerical simulations
\citep[e.g.,][]{crumley19} show substantial shock-velocity dependence
in $\epsilon_e$, as well as dependence on other shock parameters.  In
SNR studies, more recent work has attempted to simulate the
acceleration of electrons and amplification of magnetic field based on
analytic prescriptions, so that the values of $\epsilon_e$ and
$\epsilon_B$ can evolve \citep[e.g.,][]{sarbadhicary17,
  pavlovic18}. However, these studies also assume parameters such as a
fixed ratio of cosmic-ray electron to ion energy, or a fixed injection
efficiency into the acceleration process, that affect the behavior of
the $\epsilon$ factors.

It is the goal of this paper to determine as well as possible the
values of $\epsilon_e$ and $\epsilon_B$ in young supernova remnants,
using the best observational data available, and making minimal
theoretical assumptions.  We discuss methods for determining
magnetic-field strengths in Section~\ref{magfield}, and for extracting
relativistic-electron energy densities in Section~\ref{els}.  We then
apply these to estimate average $\epsilon$ factors for several young
SNRs, in Section~\ref{avg}.  The core of the paper is in
Sections~\ref{kepler} and~\ref{results}, where we obtain spatially
resolved values of the $\epsilon$ factors at seven locations around
the rim of Kepler's SNR.  The results are discussed in
Section~\ref{disc} and summarized in Section~\ref{concls}.

\section{Magnetic-field determinations}
\label{magfield}

There are various methods for obtaining independent measures of
magnetic field in compact synchrotron sources.  One interesting
method, applicable to radio supernovae, produces some valuable
information on the values and evolution of $u_B$ and $\epsilon_B$, so
shall be described here.  While this method cannot be applied to SNRs,
it does allow inferences of those quantities over time as a supernova
evolves.  The unsettling results (at least for SN 1993J) set the stage
for our considerations of SNRs.

\subsection{Synchrotron self-absorption in supernovae}

This method, applicable whenever the process operates, relies on
synchrotron self-absorption (SSA): the observation of the frequency at
which a source becomes optically thin to synchrotron radiation, and
the flux at that frequency, allow the determination of two out of the
three quantities: source size, magnetic field, and electron energy
density. However, the operation of SSA at observable radio frequencies
requires conditions in a diffuse source that are fairly restrictive.
Using the notation of \cite{pacholczyk70}, the absorption coefficient
for a homogeneous synchrotron source with magnetic field $B$ and
electron energy spectrum $KE^{-s}$ can be written
\begin{equation}
  \kappa_\nu = c_6(s) \left(1.25 \times 10^{19}\right)^{(s+4)/2} c_9(s + 1)
  K B^{(s + 2)/2} \nu^{-(s + 4)/2} \ \ {\rm cm}^{-1}
\end{equation}
where the numerical constants are given in \cite{pacholczyk70}.  For a
typical electron energy index $s = 2.5$ (synchrotron spectral index
$\alpha \equiv (s - 1)/2 = 0.75$, with $S_\nu \propto \nu^{-\alpha}$), this is
\begin{equation}
  \kappa_\nu = 6.13 \times 10^{21} K B^{9/4} \nu^{-13/4} \ \ {\rm cm}^{-1}.
\end{equation}
If $s > 2$, the energy density in electrons $u_e$ depends only on the
lower energy limit to the spectrum $E_l$.  Synchrotron emission basically
requires $E_l \gapprox 10 m_ec^2$.  An estimate of the required conditions
for observable synchrotron self-absorption can be made by characterizing
both $K$ and $B$ in terms of energy densities.  If $u_e$ and
$u_B$ are both equal to some nonthermal energy
density $u_{\rm nonth}$, then a source of line-of-sight extent $L$ will
become opaque to SSA below a frequency of about
\begin{equation}
  \nu_1 \sim 2 \times 10^6 L^{4/13} u_{\rm nonth}^{17/26} \ \ {\rm Hz}
\end{equation}
(still for $s = 2.5$), and for that frequency to exceed 100 MHz,
the source extent must satisfy
\begin{equation}
  L \gapprox 3 \times 10^5 u_{\rm nonth}^{-17/8}
  \ \ {\rm cm}.
\end{equation}
For a source extent less than 1 pc, the nonthermal energy density
must exceed about $10^{-6}$ erg cm$^{-3}$, or about 1 MeV cm$^{-3}$.
Thus SSA is an important effect only for very high energy-density
circumstances, such as supernovae -- but not SNRs.

Most radio supernovae show evolving spectra with a peak at some
frequency which moves lower with time, attributed to synchrotron
emission with some opacity setting in at lower frequencies.  Some
combination of free-free absorption, either coincident or foreground,
and SSA is likely responsible \citep[e.g.,][]{chevalier84,weiler02}.  An
extensive study by \cite{chevalier98} attributes low-frequency
absorption in 8 of 13 radio supernovae to SSA.  He reduces the three
required source parameters to two by assuming a fixed ratio (not
necessarily one) of $u_e$ to $u_B$, and derives source sizes and
magnetic fields on the assumption that $u_B \propto \rho v_s^2$, the
post-shock pressure.  Magnetic-field strengths depend fairly weakly on
all parameters, and are in the range 0.1 -- 0.6 G, at the time of the
emission peak.

In one case, SN 1993J, the radio observations are sufficiently
frequent and the frequency coverage so extensive as to allow a
detailed determination of $u_e$ and $u_B$ independently and as a
function of time \citep{fb98}, since the VLBI
observations of a more-or-less constant expansion velocity (for the
first $\sim 100$ days) of $2 \times 10^4$ km s$^{-1}$ \citep{bartel94}
allow the radius to be inferred.  They find that $u_e \propto \rho
v_s^2$ describes the data well, and much better than $u_e \propto
\rho$ alone (here $\rho$ is the upstream, i.e., circumstellar-medium
    [CSM], density).  They determine $\epsilon_e \sim 5 \times
    10^{-4}$.  For magnetic field, the apparent deceleration beginning
    around day 100, with $R \propto t^m$ and $m = 0.74$, allows a
    discrimination between a model with $u_B \propto \rho v_s^2$ and
    one with $B \propto 1/R$, with the latter description providing a
    better fit to data.  Before Day 100, they find $u_B/\rho v_s^2
    \sim 0.14$, independent of time.  The suggestions of non-constant
    $\epsilon_B$, and worse, of a change in the very dependence of
    $\epsilon_B$ on SN parameters, are worrisome hints that the
    simple picture of constant epsilons is a poor description of
    the processes of shock acceleration and magnetic-field amplification.
    While we can only observe SNRs evolve through a small fraction of
    their lifetimes, the results of SN 1993J should put us on notice
    that results for SNRs may fail to tell a complete picture.

The detailed SN inferences rely on a simple one-zone emission model
(though with spectral sophistication; \cite{fb98} evolve the electron
distribution under both Coulomb and synchrotron losses), and on the
operation of SSA as an absorption mechanism.  They also apply to very
high shock velocities and dense CSM.  The importance of the processes
of magnetic-field amplification and particle acceleration is
sufficiently great that testing assumptions such as the scaling of
$u_e$ and $u_B$ with density and shock velocity, and possible
evolution of efficiencies with time, in different regions of parameter
space, is a high priority.  It is fortunate that methods exist to
allow this in SNRs, especially to replace SSA for magnetic-field
determinations.

\subsection{Magnetic-field determinations in supernova remnants}

SNRs are far too diffuse for SSA to be an important mechanism, so
inferences of magnetic-field strengths rely on different techniques.
\cite{reynolds12} review these.  Of particular interest is the ``thin
rims'' argument, based on observations of X-ray synchrotron emission
from young SNRs \citep{bamba03,vink03,parizot06}, in particular, on
the commonly observed morphology of thin tangential rims at the shock
front (located, for instance, by H$\alpha$ observations).  This method
relies on the assumption that the disappearance of emission a short
distance downstream results from synchrotron losses on the emitting
electrons as they are advected (or diffuse) downstream.  A thorough
treatment is given by \cite{ressler14}, who also include a discussion
of an alternative explanation for thin rims, decay of magnetic
turbulence downstream \citep{pohl05}.  Under the simplest
assumptions (electron transport by pure advection, ignoring
magnetic-field damping, the delta-function approximation for the
single-electron emitted spectrum), a straightforward relation can be
derived between rim thickness and magnetic-field strength
\citep{parizot06}:
\begin{equation}
  B \cong  210 \left( \frac{v_s}{1000 \ {\rm km\ s}^{-1}} \right)^{2/3}
  \left( \frac{w}{0.01\ {\rm pc}} \right)^{-2/3} \ \mu{\rm G}
  \label{rimfield}
  \end{equation}
where $w$ is the filament width in the radial direction.  (We have
assumed a small geometric correction factor ($4 \bar{P}/r_{\rm comp}$) to be
unity, where $r_{\rm comp}$ is the compression ratio and $\bar{P} = 1$ for a
perfect sphere.  Since we will always be employing this relation in
small regions at the very edge of the remnant, the locally spherical
approximation should be quite good.)  However, more elaborate
treatments produce somewhat different values; \cite{ressler14} collect
published values for the remnant SN 1006 ranging from 65 to 130
$\mu$G, with one report of 14 $\mu$G.  All agree, however, in
requiring amplification of magnetic field beyond simple shock
compression.  The situation is made more complicated by the presence
in a few cases of thin radio rims, produced by electrons with energies
far too low to be affected by synchrotron losses, and thus requiring
magnetic damping, a process with few theoretical or observational
constraints.

Given $B$, we then have the magnetic-field energy density
$u_B \equiv B^2/8\pi$.  We note that if $B$ is determined in this
way, a prediction results for $\epsilon_B \propto B^2/v_s^2 \propto
v_s^{-2/3}$, other things being equal.  We shall apply Equation~\ref{rimfield}
in the analysis below.


\section{Relativistic-electron energy densities}
\label{els}

We shall use observations of radio synchrotron intensity to obtain
relativistic-electron energy densities, assuming simple power-law
electron energy distributions.  We take $N(E) = KE^{-s}$ electrons
cm$^{-3}$ erg$^{-1}$.  The electron energy density is then
\begin{equation}
  u_e \equiv \int_{E_l}^{E_h} K\,E^{1 - s} dE = \frac{K}{s - 2}E_l^{2-s}
  \left[1 - \left(\frac{E_h}{E_l}\right)^{2-s}\right].
\end{equation}
The energy ranges are somewhat arbitrary; for true synchrotron
emission, $E_l \sim 10 m_ec^2$ is a reasonable estimate, while
since we shall have $s > 2$ and $E_h \gg E_l$, $u_e$ is essentially
independent of the value of $E_h$.  So we approximate
\begin{equation}
  u_e = \frac{K}{s-2}E_l^{2 - s}.
  \label{ue0}
  \end{equation}

Our strategy for finding $K$ and hence $u_e$ will be to observe
synchrotron intensities, which in the optically thin limit are
just proportional to the synchrotron emissivity $j_\nu$.
The synchrotron emissivity can be written (again in the notation of
\cite{pacholczyk70})
\begin{equation}
  j_\nu = c_5 (2c_1)^{\alpha} KB^{(s+1)/2} \nu^{-\alpha} \ {\rm erg\ cm}^{-3}
  \ {\rm s}^{-1} \ {\rm Hz}^{-1} \ {\rm sr}^{-1}.
 \label{jnu}
\end{equation}
Here $c_5$ is a function of the electron index $s$: $c_5 \sim
10^{-24}$ cgs for $2 < s < 3$, and $c_1 \equiv 6.27 \times 10^{18}$
cgs.  The constants can be conveniently lumped together as
\begin{equation}
c_j \equiv c_5 (2c_1)^\alpha.
\end{equation}
For $s = 2.42$, its value for Kepler, $c_j(2.42) \equiv c_5(2.42)
(2c_1)^{0.71} = 3.69 \times 10^{-10}$ cgs.  Then
\begin{equation}
  K = c_j^{-1} B^{-(s + 1)/2} \nu^\alpha j_\nu.
\end{equation}
We neglect synchrotron self-absorption for SNRs, so the intensity on
any line of sight is just given by $\int j_\nu dl$.

We can infer global average values of the emissivity by measuring the
total remnant flux at frequency $\nu$, $S_\nu$.  For spatially
resolved measurements, we shall measure local values of $I_\nu$ and
estimate (short) line-of-sight depths $L$ so $j_\nu = \langle I
\rangle/L$.  For a homogeneous source of volume $V_{\rm em}$ at
distance $d$, we can write (assuming isotropic synchrotron emission,
that is, disordered magnetic field)
\begin{equation}
  S_\nu = (4\pi j_\nu) \frac{V_{\rm em}}{4\pi d^2}
  \label{Snu}.  
\end{equation}
%
We will first apply this to obtain a mean $u_e$ for a spherical
remnant of radius $R$ where we take the emitting volume to be $V_{\rm
  em} = (1/r_{\rm comp})4\pi\,R^3/3$,
where $r_{\rm comp}$ is the
shock compression ratio. So
%
$j_\nu = d^2 S_\nu/V_{\rm em}.$
Finally, the electron energy density $u_e$ is given by
\begin{equation}
  u_e =\left[ E_l^{2 - s}/(s - 2) \right] c_j^{-1}\, B^{-(s + 1)/2}\, \nu^\alpha\,
    d^2\, S_\nu \left( 3r_{\rm comp} /4\pi\,R^3 \right).
  \label{ue1}
\end{equation}
All quantities on the right are observable with the exception of the assumed
value of $E_l$ of $10 m_e c^2$.

For Kepler's SNR, we shall consider small regions at the very edge
of the remnant, and shall assume that $j_\nu$ is constant along
our (short) lines of sight $L$, so the intensity $I_\nu = j_\nu L$.
Then we obtain an expression for the electron energy density $u_e$:
\begin{equation}
  u_e = \left[c_j(s) (s - 2)\right]^{-1} E_l^{2 - s} I_\nu L^{-1}
  B^{-(s+1)/2}\nu^{\alpha}.
  \label{ue2}
  \end{equation}
Finally, we obtain $\epsilon_e \equiv u_e/\rho_0 v_s^2$ in either case.

\section{Spatially integrated determinations}
\label{avg}

Here we present spatially-averaged determinations of $u_e$ and $u_B$
for several young remnants: G1.9+0.3 (age $\sim 100$ yr); Cas A ($\sim
350$ yr); Kepler (417 yr); Tycho (449 yr); SN 1006 (1015 yr); and RCW 86
\citep[1836 yr, if SN 185 CE;][]{williams11b}.  We collect observations
from the literature.  Table~\ref{rimfieldtable} lists magnetic-field
determinations using variants of Eq.~\ref{rimfield} for these objects.
The large dispersion in values from different authors illustrates the
assumption-dependence of the analysis.  The most complete discussion,
including many subtle effects, can be found in \cite{ressler14}.
Additional data required to infer the mean relativistic-electron
density are given in Table~\ref{global}.  Averaging over the entire
remnant is a crude application of the magnetic-field results that
strictly apply only to the particular rims which were measured.

For the results in Table~\ref{global}, we chose the magnetic-field
values of \cite{parizot06}, except for RCW 86 for which we used the
value from \cite{volk05}.  The value for G1.9+0.3 was obtained from
the width of a thin rim in the west edge of $1.8''$ measured from the
2011 image \citep{borkowski13}, and using $v_s = 14,000$ km s$^{-1}$
and $d = 8.5$ kpc \citep{reynolds08}.  The various values in
Table~\ref{rimfieldtable} range over about a factor of 3-4 at most,
introducing a possible range of an order of magnitude in the values of
$u_B$, $u_e/u_B$, and $\epsilon_B$ in Table~\ref{global}.  Even with
this large possible range, however, the spread in values of $u_e/u_B$
(independent of pressure) and $\epsilon_B$ is far greater.  Given the
heterogeneous input data, we have not attempted to quantify
uncertainties, but it is likely that those due to the magnetic field
are the dominant contribution to the error budget.  At any rate, it
seems exceptionally unlikely that the large spread in quantities such
as $u_e/u_B$ could be due to measurement uncertainties.

\begin{deluxetable*}{llllll}
  \tablecolumns{6}
  \tablecaption{Magnetic Field Strengths in Young Remnants\label{rimfieldtable}}
\tablehead{
\colhead {Object} & P+06 & VBK05 & RP12 & RP12 & T+15\\
&&&Loss &Damp & }
\startdata
Cas A          &210-230  &500    & 520 & 115-260 &\\
Kepler         &170-180  &200    & 250 & 80-135  &\\
Tycho          &200-230  &300    & 310 & 85-150  & 50-400\\
SN 1006        &57-90    &140    & 130 & 64-65   & 40-200\\
RCW 86         &         &100    &     &         &       \\
\enddata
\tablecomments{Magnetic-field strengths are in $\mu$Gauss.  References:  P+06, \cite{parizot06}; 
VBK05, \cite{volk05}; 
RP12, \cite{rettig12}; 
T+15, \cite{tran15}}   
\end{deluxetable*}

\begin{deluxetable*}{lcccccccccccccc}
  \tablecolumns{8}
  \tablecaption{Derived Properties of Remnants\label{global}}
  \tablehead{
    \colhead {Remnant} & $B$ & $S_9$ & $d$ & $\alpha$\tablenotemark{a} & $R$ & $u_e$\tablenotemark{b} & $u_B$\tablenotemark{b} & $u_e/u_B$ & $n_0$ & $v_{\rm sh}$ & $p$\tablenotemark{c} & $\epsilon_e$ & $\epsilon_B$ & References\\
    & ($\mu$G) & (Jy) & (kpc) & & (arcsec) & & & & (cm$^{-3}$) & (km s$^{-1}$) & & ($10^{-3}$) & ($10^{-3}$) &}
  \startdata
  G1.9+0.3 & 320\tablenotemark{d} &  0.6 & 8.5 & 0.6  &   50 & 0.22  & 41 & 0.0055   & 0.02 & 14,000 & 0.91  & 0.25   & 45  & 1 \\
  Cas A    & 220                  & 2300 & 3.3 & 0.77 &  150 & 290   & 19  & 15      & 1    &  5800  & 3.3   & 86     & 5.8 & 2 \\  
  Kepler   & 175                  &   18 & 5   & 0.71 &  117 & 3.6   & 12  & 0.30    & 3    &  4000  & 11    & 0.58   & 1.1 & 3 \\
  Tycho    & 215                  &   50 & 2.4 & 0.6  &  240 & 1.1   & 18  & 0.060   & 0.2  &  2000  & 0.19  & 5.9    & 100 & 4 \\
  SN 1006  &  74                  &   19 & 2.2 & 0.6  &  900 & 0.050 & 2.1 & 0.023   & 0.05 &  5000  & 0.29  & 0.17   & 7.4 & 5 \\
  RCW 86   & 100\tablenotemark{e} &   49 & 2.5 & 0.6  & 1260 & 0.025 & 4.0 & 0.0063  & 0.5  &   600  & 0.042 & 0.60   & 96  & 6 \\
    \enddata
    \tablecomments{Magnetic field values from \cite{parizot06} except as noted.
      References: 1, \cite{carlton11}; 2, \cite{chevalier11}; 3, \cite{williams12}; 4,
      \cite{williams13}; 5, \cite{winkler14}; 6, \cite{williams11b}}
    \tablenotetext{a}{From \cite{green19}.}
    \tablenotetext{b}{In units of $10^{-10}$ erg cm$^{-3}$.}
    \tablenotetext{c}{Pressure ($\rho v_{\rm sh}^2$) in units of $10^{-7}$ dyn cm$^{-2}$.}
    \tablenotetext{d}{Obtained from a new measurement.  See text.}
    \tablenotetext{e}{Obtained from \cite{volk05}.}
\end{deluxetable*}

\section{Spatially resolved efficiencies in Kepler's supernova remnant}
\label{kepler}

The surprisingly large range of values of the various efficiency
parameters listed in Table~\ref{global} applies across rather
different SNRs, of different ages and properties.  The environments
  of these SNRs are likely very different as well; one might hope to
  attribute the results of Table~\ref{global} to this heterogeneity.
However, it is
possible to use similar methods to analyze the conditions at different
locations around the rim of a single remnant.  This technique has
obvious benefits in removing dependence on many quantities that vary
between remnants, chiefly but not exclusively distance.  Here we
describe a spatially resolved analysis of the efficiencies of electron
acceleration and magnetic-field amplification at different locations
around the periphery of Kepler's supernova remnant (``Kepler''
hereafter).  The distance to Kepler is uncertain, with values ranging
from 4 kpc \citep{sankrit05} to 8 kpc
\citep[e.g.,][]{millard19}. Recent optical determinations suggest 5
kpc \citep{sankrit16}, and we shall assume 5 kpc here.  However, the
main import of our results will be in comparing inferences at
different points, hence independent of distance.

\subsection{Observational strategy}

To fix the efficiencies of shock-energy deposition into magnetic field
and relativistic electrons, one requires independent local measurements of
density, shock velocity, magnetic-field strength, and electron energy
density.  Such measurements have substantial uncertainties for a SNR.  X-ray
diagnostics of plasma density are fraught with uncertainties,
requiring spectral modeling and assumptions about filling factors.
Shock velocity determinations from X-ray spectra suffer from
uncertainty about prompt electron heating in the shock.  These
difficulties require alternative methods for determining each
quantity.

An excellent diagnostic for the density of a hot plasma is the
temperature of grains embedded in it \citep[e.g.,][]{dwek92}, which
are collisionally heated by the plasma.  We have previously employed
this method to obtain post-shock densities of SNRs in the Large
Magellanic Cloud \citep{borkowski06,williams06}, and of Kepler
\citep{williams12}.  The latter study involved analysis of a complete
spectral mapping of Kepler with the {\sl Spitzer} IRS, and extraction
of spectra between 7.5 $\mu$m and 38 $\mu$m at several locations.
These spectra were fit with a dust-heating shock model in which
post-shock grains were heated as they were advected
downstream in a constant-temperature, constant-density plasma
\citep{borkowski06}.  This model provided good descriptions of
observed IR spectra, allowing extraction of post-shock densities at various
points in Kepler.  These densities are listed in Table~\ref{datatable}.
We require $\rho_0$, which we infer by dividing those values by an
assumed compression ratio $r_{\rm comp}$ of 4, and assuming cosmic
abundances (mean mass per hydrogen atom $\mu = 1.4$).

For shock velocities, we use proper motions of expansion, as measured
from {\sl Chandra} X-ray data \citep{katsuda08}, in 14 regions around
the periphery.  We then deduce magnetic-field strengths using the
simple thin-rims analysis of Eq.~\ref{rimfield}, measured from the
741-ks {\sl Chandra} image \citep{reynolds07}, the same image used for
the second epoch of proper-motion measurements \citep{katsuda08}.
Then the electron energy density can be extracted from intensities
measured from a radio synchrotron image \citep[5 GHz
  VLA;][]{delaney02} using Eq.~\ref{ue2}.

Kepler's SNR is an excellent target with which to attempt this
determination.  The very strong N-S brightness gradient at all
wavelengths indicates that the shock wave is expanding into highly
asymmetric material; \cite{blair07} find that the density to the north
is 4 -- 9 times that to the south (see also \cite{williams12}), and
the shock velocities show up to a factor 3 of variation
\citep{katsuda08}.  These wide variations allow us to examine the
dependence of efficiencies on several factors.  Thin X-ray rims with
nonthermal spectra can be seen in several locations around the
periphery. Below we outline the quantitative results of our
investigation.

\begin{figure}
  \centerline{
    \includegraphics[width=3truein]{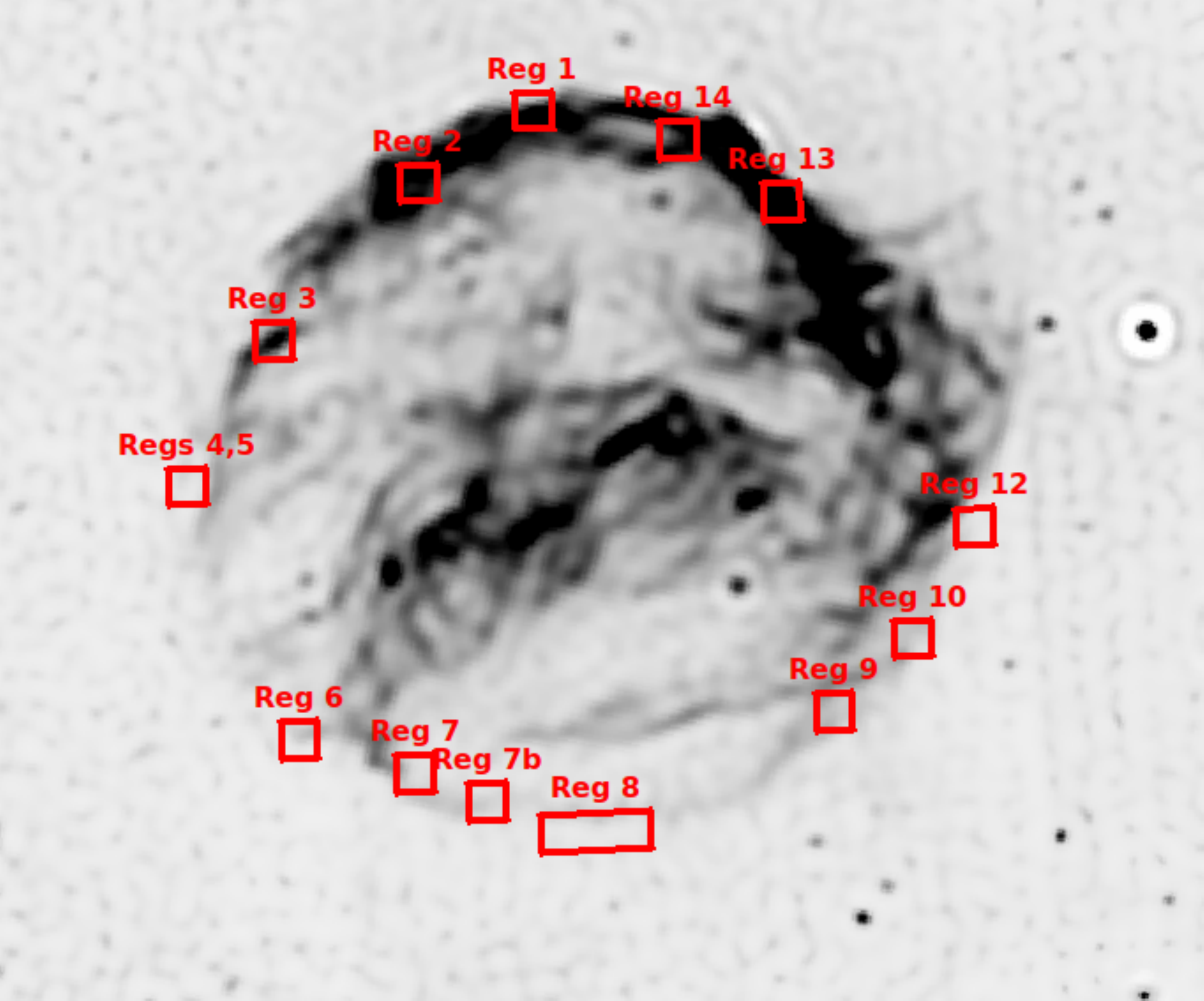}\hskip0.1truein
    \includegraphics[width=3truein]{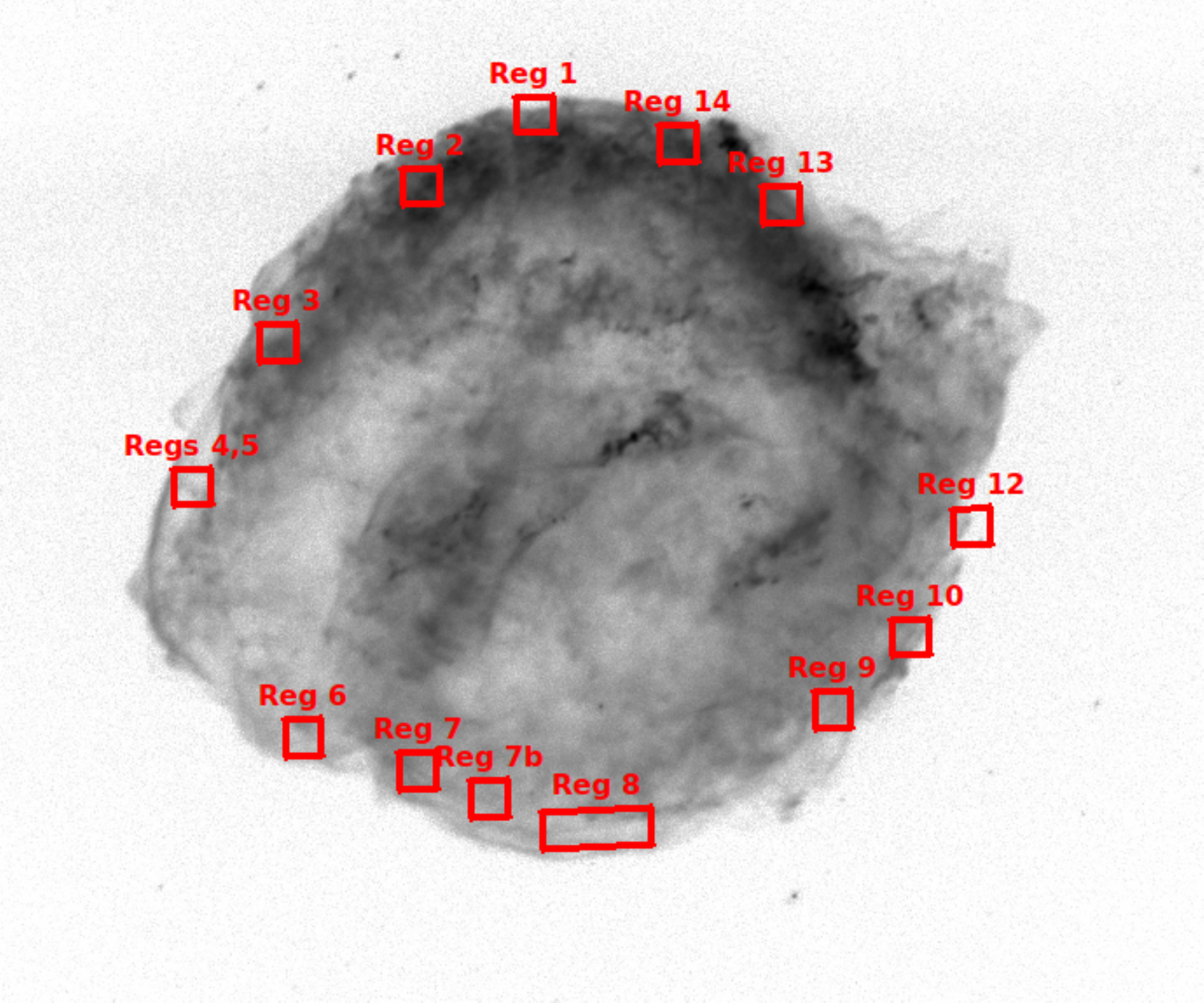}}
\caption{Left: Deconvolved {\sl Spitzer} 24 $\mu$m image
  \citep{williams12}, with regions superposed.  Right: {\sl Chandra}
  image, 0.3 -- 7 keV \citep{reynolds07}.}
  \label{ir-x-regs}
\end{figure}  

\begin{figure}
  \centerline{
    \includegraphics[width=3truein]{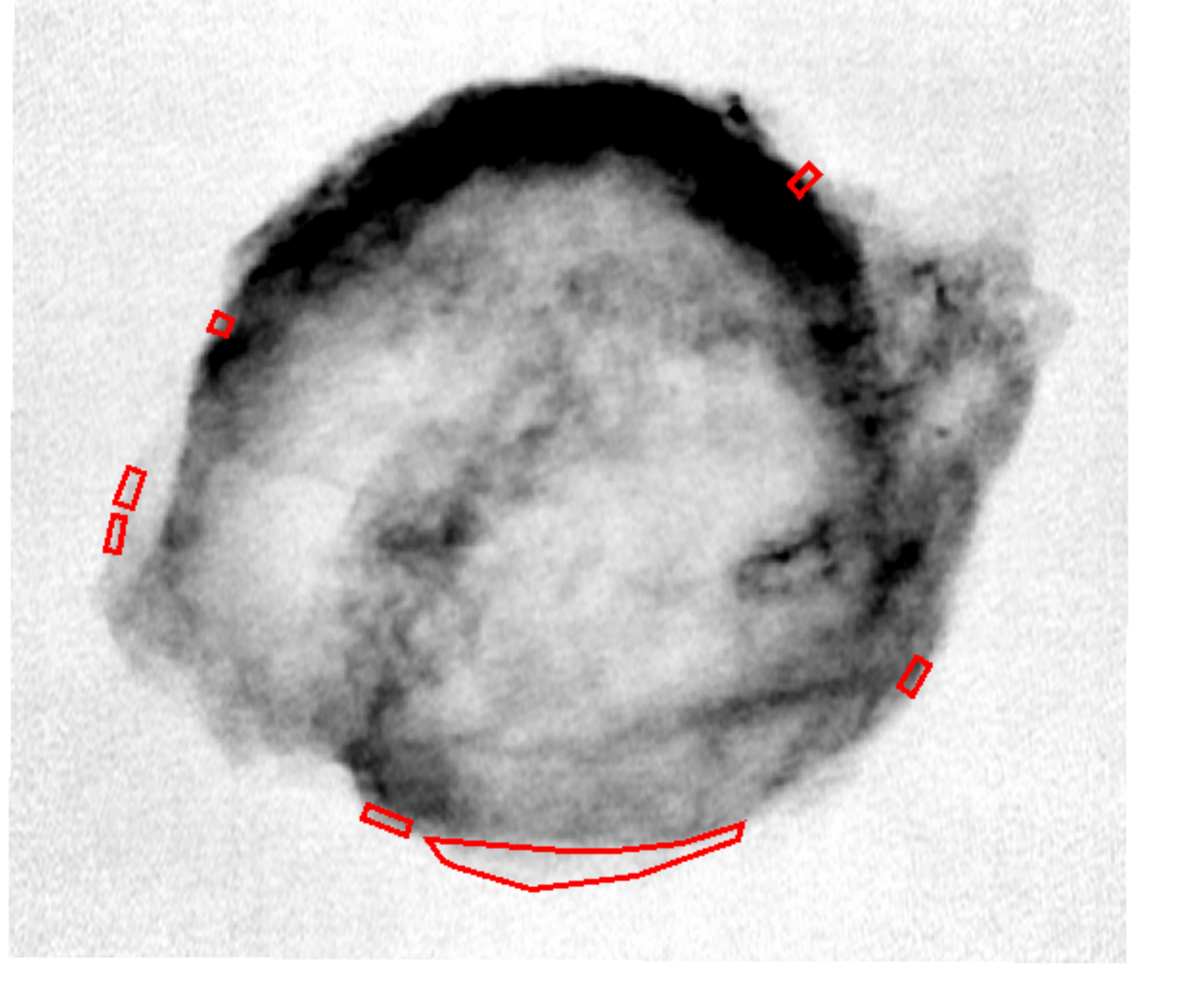}\hskip0.1truein
    \includegraphics[width=3truein]{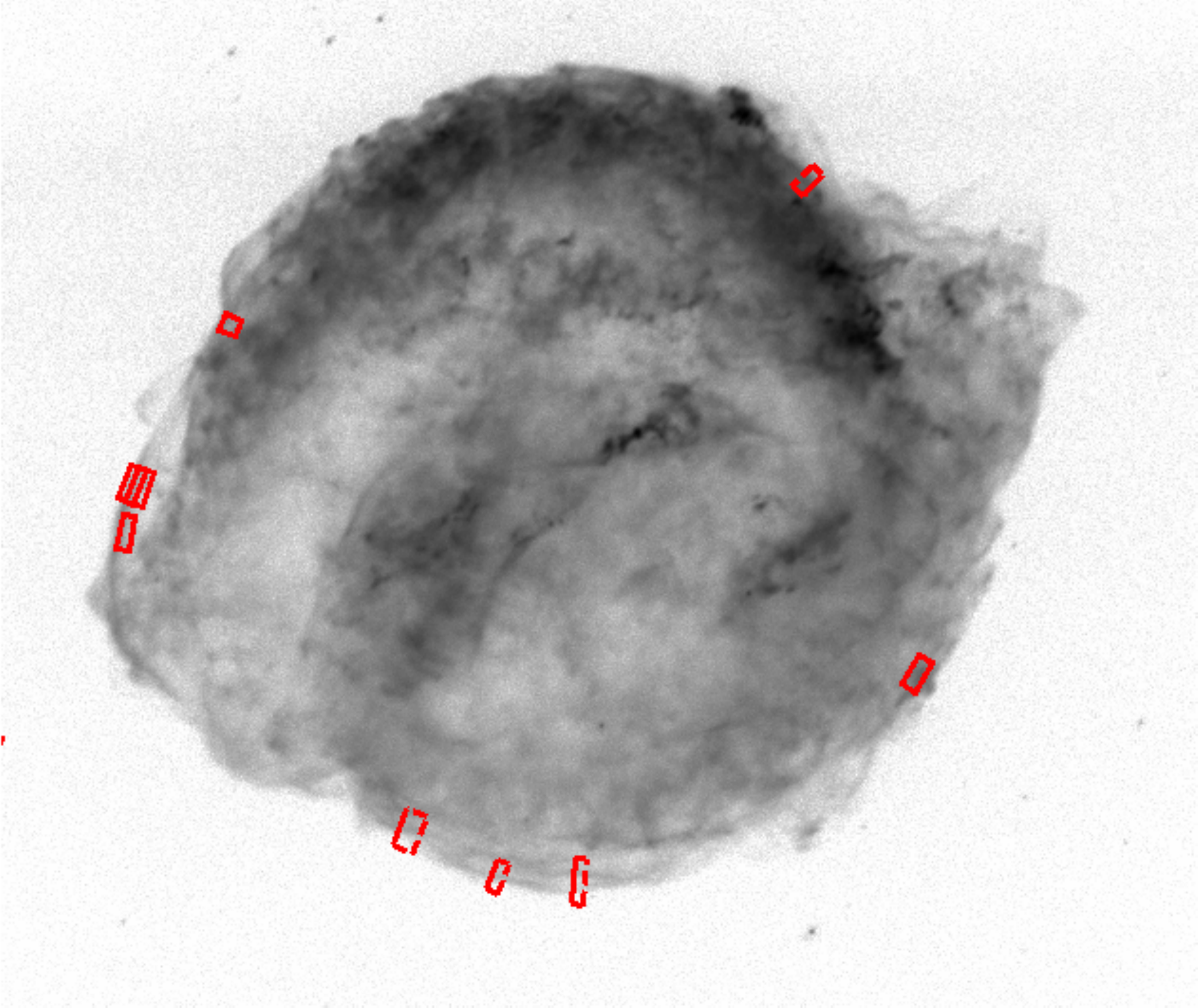}}
\caption{Left: VLA 5 GHz image \citep[epoch 1997;][]{delaney02}, showing
  regions used for measurements of radio intensity.
These regions approximate the locations of
  \cite{katsuda08} where shock velocities were measured, and the
  regions of Fig.~\ref{ir-x-regs} where densities were obtained.
  Right: {\sl Chandra} image (2006), where regions shown are where filament
  widths were measured.  Region 4 is shown in two locations; the slightly
  more interior one is identical to that on the radio image to the left,
  but expansion between 1997 and 2006 caused the thin filament to move
  beyond the original region.  The filament morphology did not
  change appreciably between 2000 and 2006, as shown in the difference
  image of \cite{katsuda08}, so the slightly expanded outer region was
  used to measure the profile.
  \citep{reynolds07}.} 
\label{rxregions}

\end{figure}  

\subsection{Density measurements}

\begin{figure}
  \centerline{\includegraphics[width=4truein]{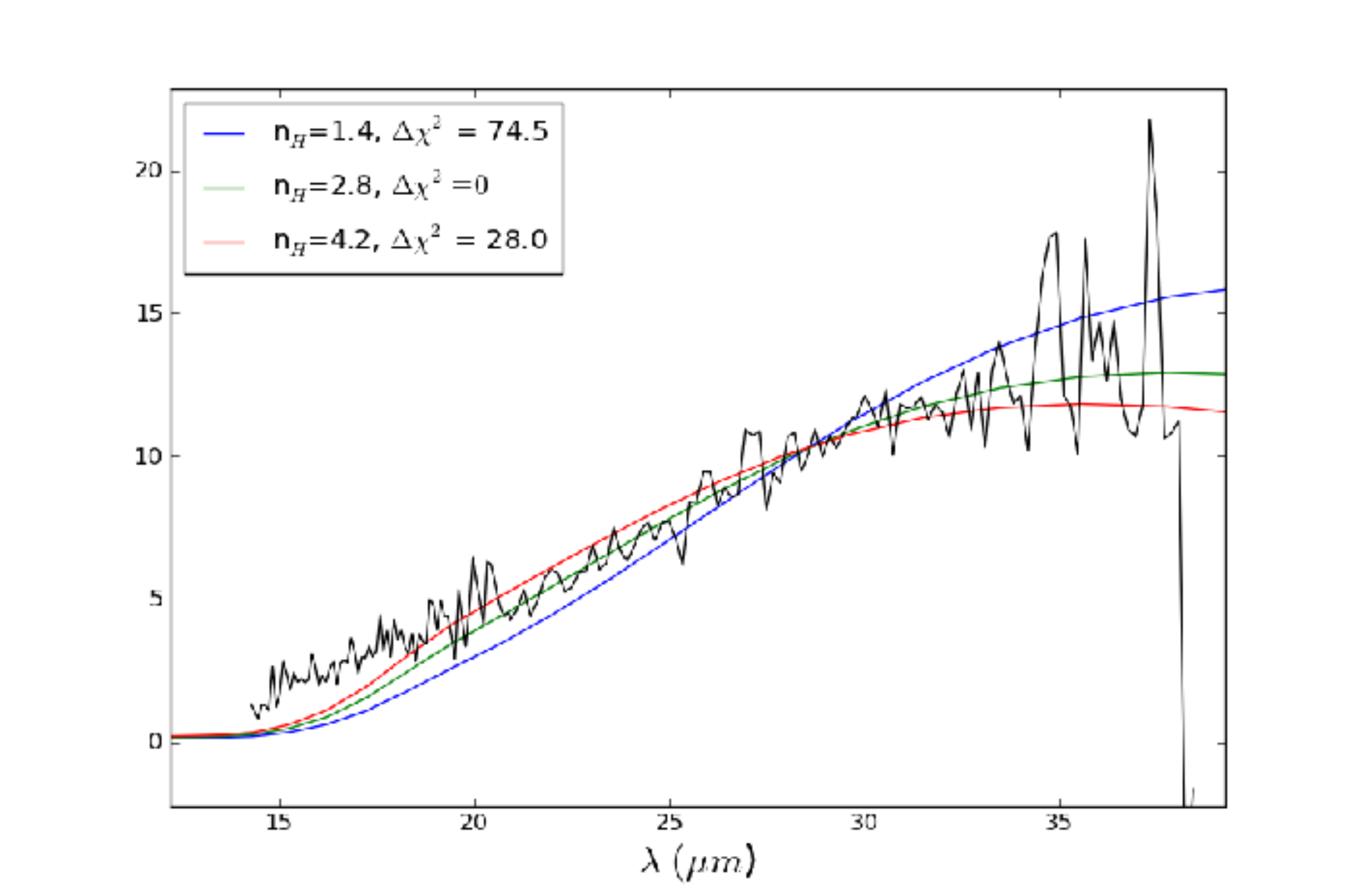}}
  \caption{Region 10 observed {\sl Spitzer} LL spectrum, with three
    models shown.  See \cite{williams12} for details.  
    The ion temperature is set by the observed shock
    velocity and adiabatic shock jump conditions; the electron
    temperature is assumed to be 1.5 keV for all regions.  A
    least-squares algorithm is used to fit models to data, fitting
    between 21 and 33 $\mu$m; shorter-wavelength emission reflects
    smaller grains and greater model uncertainty.}
  \label{3dens}
  \end{figure}

The determinations of post-shock density were based on observations of Kepler
with the {\sl Spitzer} IRS instruments: both orders of the
low-resolution (LL) module (14 -- 38 $\mu$m), and order 1 (7.5 -- 14
$\mu$m) of the short-wavelength low-resolution (SL) module.  The
entire remnant was mapped with the LL instrument, and selected regions
with the SL module.  The observations and analysis are described in
\cite{williams12}.
Figure~\ref{3dens} illustrates the sensitivity of model spectra to gas
density.  Models depend weakly on both ion temperature $T_i$ and
electron temperature $T_e$ (see \cite{blair07} and \cite{williams12}
for details).
Observed electron temperatures in young SNRs are typically a few keV
\citep{reynolds07}; we assume $kT_e = 1.5$ keV, but set uncertainties
by allowing a range between 1 and 2 keV.

Broadly, variations of a factor of 30 were found in density between
the faint southern rim and bright north, but values were obtained at
many regions around the periphery.  (Averaging over larger regions
reduces the contrast to the range of 4 -- 9 quoted in \cite{blair07}.)
Figure~\ref{ir-x-regs} shows regions from which spectra were
extracted, models fit, and densities obtained.  Those densities are
tabulated in Table~\ref{obs}.  Statistical uncertainties of the fits
are much smaller than the spread of values possible by changing the
assumed electron temperature between 1 and 2 keV; that spread is shown
in Figure~\ref{nh_vs-pr}.

While we believe the emission from each region is well-characterized
by the densities listed in Table~\ref{obs}, in a few regions the IR
emission is so faint that it is not clear if the densities listed
there describe the immediate post-shock density.  Fig.~\ref{ir-x-regs}
shows that for Regions 4 and 5, 8, and 12, detectable IR is only at
the innermost edge.  Both Figs.~\ref{ir-x-regs} and~\ref{rxregions}
show the location of the blast wave as indicated by radio and
nonthermal X-rays; the absence of immediate IR emission there suggests
density variations along the line of sight, with IR appearing only
once the density is somewhat larger (and shocks somewhat slower).  For
those regions, we have chosen to regard the densities of
Fig.~\ref{obs} as upper limits.  Higher spatial-resolution IR observations,
possible with JWST, will be required to improve our knowledge of the
immediate post-shock density in all locations.

\subsection{Shock velocities}

\begin{figure}
  \centerline{\includegraphics[width=2.5truein]{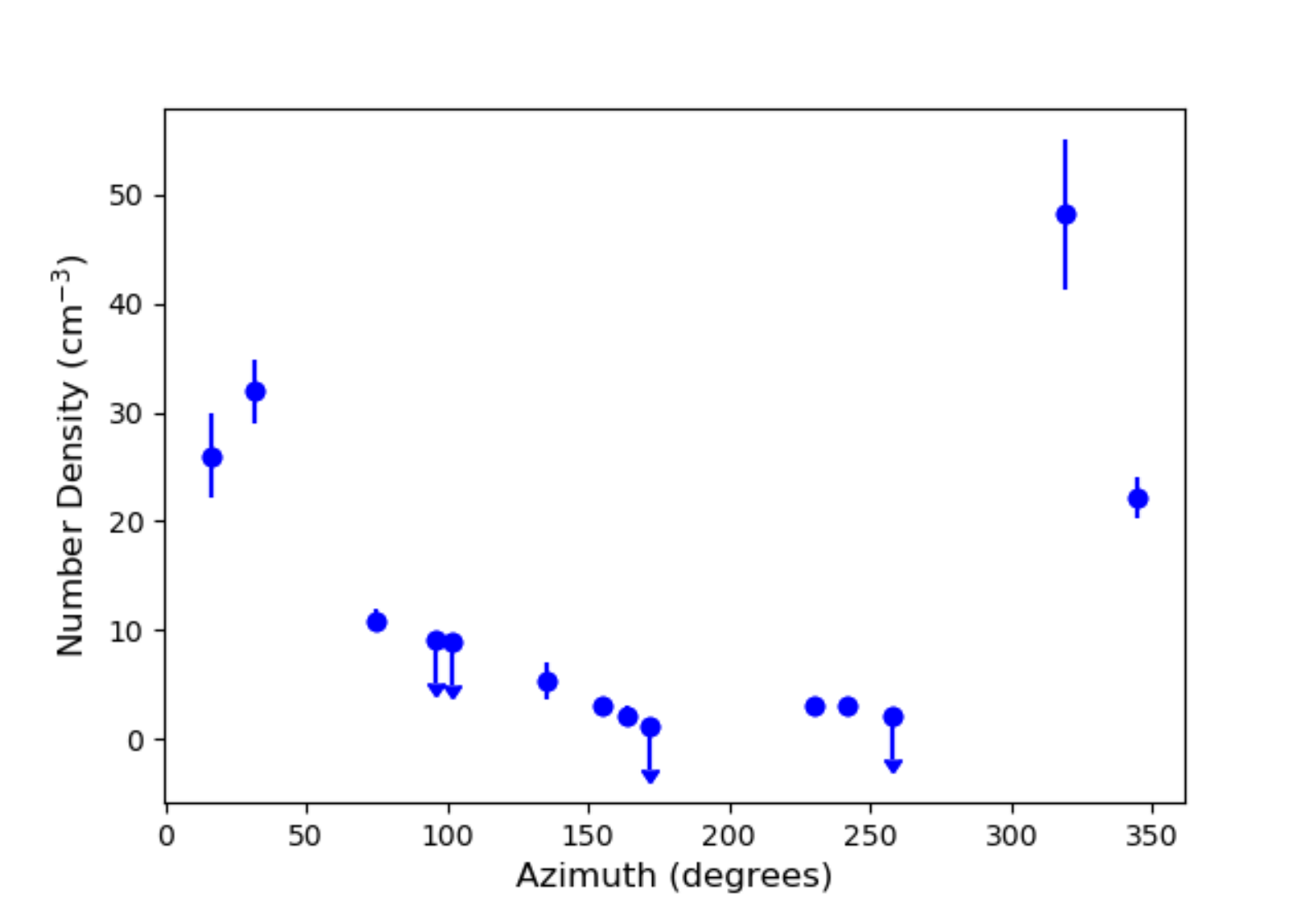} \hskip0.1truein
    \includegraphics[width=2.5truein]{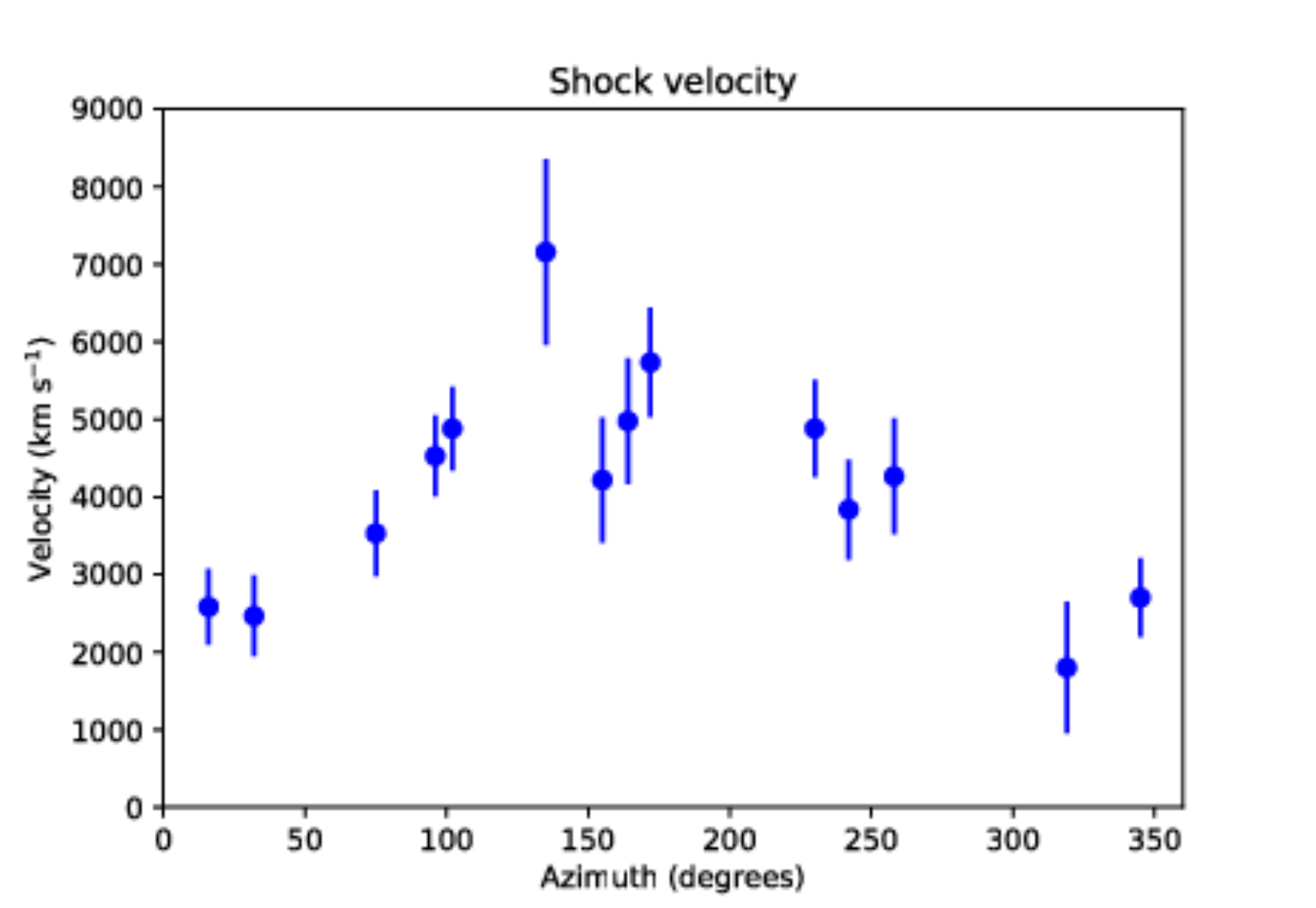} \hskip0.1truein
    \includegraphics[width=2.5truein]{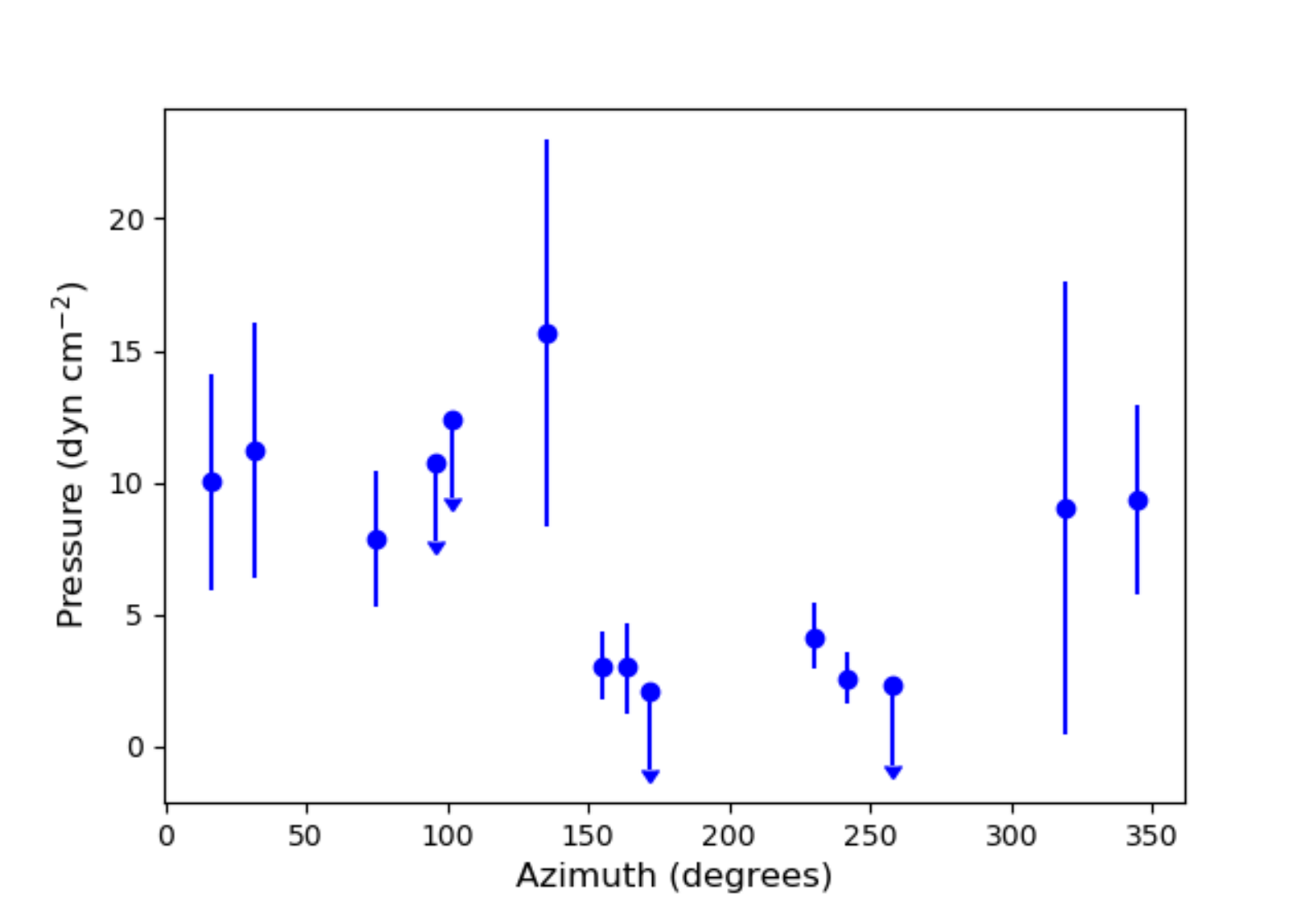}}  
  \caption{Observed post-shock densities (left) and shock velocities
    \citep[center;][]{katsuda08} at indicated positions.  Uncertainties
    in density are given by allowing the assumed electron temperature
    to vary between 1 and 2 keV.  Right: Pressure at different
    locations.  Neutral upstream gas ($\mu = 1.4$) is assumed.
    Pressures are highest in the north, where densities are largest.}
  \label{nh_vs-pr}
  \end{figure}

Proper motions of expansion were measured by \cite{katsuda08}
comparing {\sl Chandra} images obtained in 2000 and 2006, using images
between 1.0 and 8.0 keV.  We have converted them, and their
uncertainties, into velocities for our nominal distance of 5 kpc.
(Regions shown in Fig.~\ref{ir-x-regs} correspond to whole {\sl
  Spitzer} IRS pixels.  They overlap, but are not identical with, the
corresponding regions used by Katsuda et al.~2008.)
Figure~\ref{nh_vs-pr} shows both densities and shock velocities of
various regions.  As expected, these quantities anti-correlate.
Densities vary fairly smoothly around the periphery of Kepler, with lowest
values in the south and highest in the north, well correlated with
the brightness in radio or X-rays, while the shock velocity (with larger
errors) also varies fairly smoothly in the opposite sense.
Post-shock pressures $P_2 \equiv \rho v_s^2$ (again, eliding the factor
$2/(\gamma - 1)$) are shown in
Figure~\ref{nh_vs-pr}, where we have assumed $\rho_0 = 1.4 n_H\, m_H/r_{\rm comp}$,
appropriate for neutral gas upstream.  We take $r_{\rm comp} = 4$.

The pressure varies by about a factor of a few around the periphery of
Kepler, not unexpected given the very strong gradient in external
density, with highest pressure to the north, where the external
density is much larger.  This magnitude of pressure variation is
comparable to the radial pressure gradient in the interior of a Sedov
blast wave, where the dynamical timescale is comparable to the age.

For Regions 4 and 5, 8, and 12, we regard our density determinations
as upper limits, hence upper limits on the pressure.  The variations
in pressure we find, however, are relatively small
(Fig.~\ref{nh_vs-pr}), suggesting that the true densities in those
regions are unlikely to be far below those in closely neighboring
regions.

\subsection{Magnetic-field measurements}

\begin{figure}
  \centerline{\includegraphics[width=4truein]{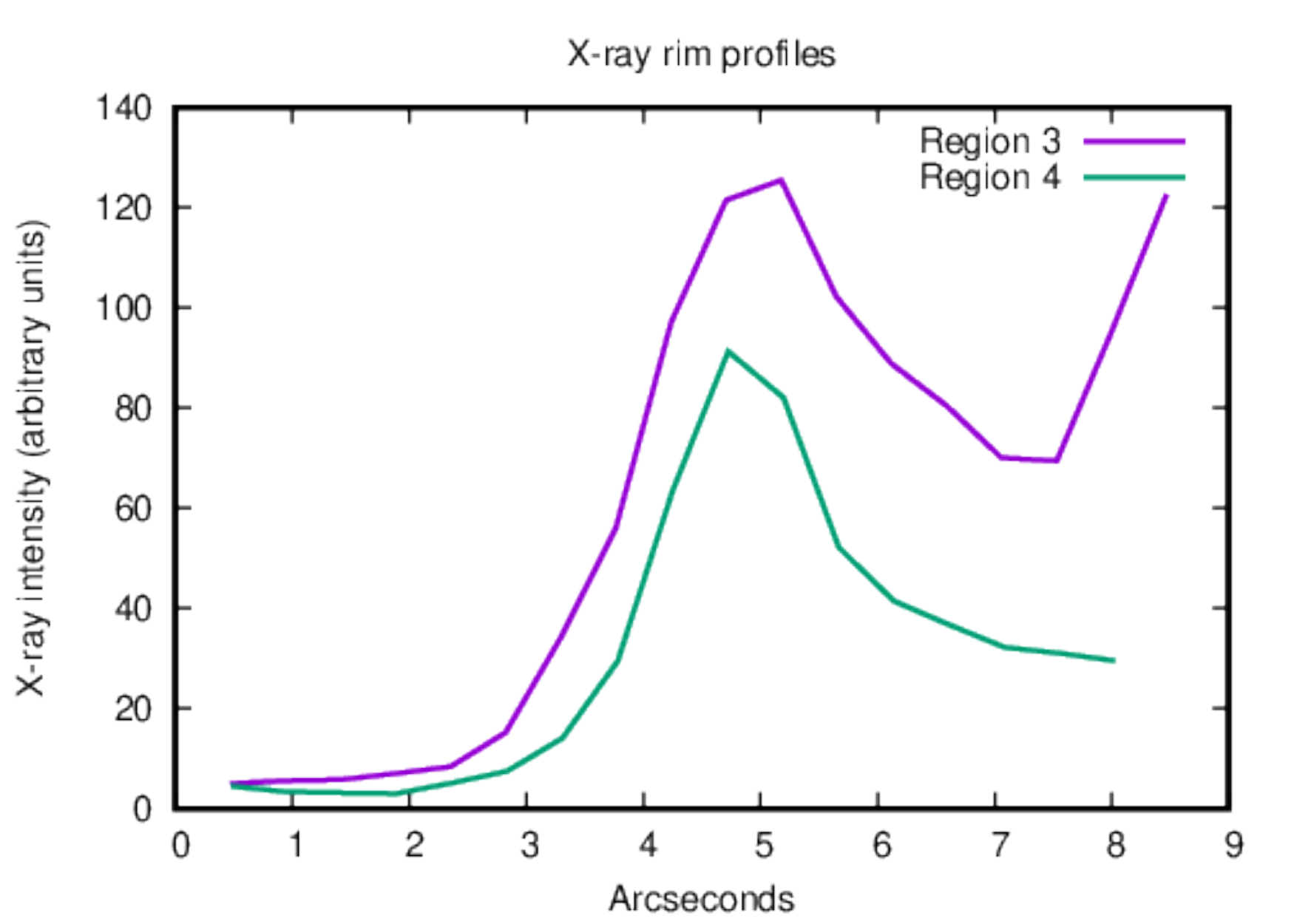}}
    \caption{X-ray rim profiles of Regions 3 and 4.  Measured
      widths are FWHMs from these figures.  Estimated uncertainties
      are $0.2''$.}
    \label{rimprofiles}
    \end{figure}

Regions shown in Fig.~\ref{rxregions} were selected for the presence
of a ``thin rim'' of nonthermal X-rays, from which a magnetic-field
strength could be extracted, making the simplest assumptions about
thin-rim physics as embodied in Equation~\ref{rimfield}. Radial
profiles (averaged azimuthally over the region's azimuthal dimensions)
were extracted; two examples are shown in Fig.~\ref{rimprofiles}.
Widths were measured at the intensity level halfway between the rim
peak and the interior minimum.  Uncertainties in this process were
estimated to be $0.2''$, although as we shall see below, the exact
value turns out not to be critical. At 5 kpc, $10^{-2}$ pc = $0.41''$,
so for a radial width $w$ of a rim in arcsec,
\begin{equation}
  B = 116 (v_{\rm sh}/10^8\ {\rm cm \ s}^{-1})^{2/3} (w'')^{-2/3} \mu{\rm G}.
\label{rimfieldsec}
\end{equation}
Note that since our shock velocities are obtained from angular
  proper motions, the ratio $v_{\rm sh}/w$ is independent of distance;
  we have also assumed 5 kpc to obtain the values for $v_{\rm sh}$ in
  Table~\ref{datatable}.  Values of magnetic field inferred from
Eq.~\ref{rimfieldsec} are listed in Table~\ref{derived}.  Not all the
rim locations of Table~\ref{datatable} showed distinct rims, so we
restrict our analysis to the seven listed in Table~\ref{derived}.
Uncertainties are difficult to gauge, but given the large spread shown
in Table~\ref{rimfieldtable}, seem to be of order a factor of 2.  The
relatively narrow range of measured filament widths results in a
(noisy) trend of inferred $B$-values with increasing shock velocity,
as expected from Eq.~\ref{rimfieldsec}.

\begin{deluxetable*}{lcccccc}
  \tablecolumns{7}
  \tablecaption{Observed Properties of Rim Regions\label{obs}}
  \tablehead{
 \colhead{Region\tablenotemark{a}} & PA & Density $n_H$\tablenotemark{b} & $v_{\rm sh}$\tablenotemark{c} & Width & Length & $\langle I_9 \rangle$ \\
    & (deg) & (cm$^{-3}$) & (km s$^{-1}$) & (arcsec) & (arcsec) & (mJy/arcsec$^{-2}$) }
  \startdata
  1  &  16 & 26 (22, 31)      & 2580 $\pm 490$ &     &    & \\
  2  &  32 & 32 (29, 39)      & 2470 $\pm 520$ &     &    & \\
  3  &  75 & 11 (10, 13)      & 3530 $\pm 550$ & 3.3 & 10 & $0.54 \pm 0.06$ \\
  4  &  96 & $<9$ (8, 10)        & 4530 $\pm 520$ & 2.3 & 33 & $0.13 \pm 0.04$ \\
  5  & 102 & $<9$ (8, 10)\tablenotemark{d} & 4880 $\pm 540$ & 2.4 & 33 & $0.17 \pm 0.05$ \\
  6  & 135 & 5.3 (3.6, 7.0)   & 7160 $\pm 1200$ &     &    &  \\
  7  & 155 & 3.0 (2.6, 3.6)   & 4220 $\pm 810$ & 4.0 & 17 & $0.35 \pm 0.04$\\
 7b\tablenotemark{e} & 164    & 2.1 (1.1, 3.1) & 4980 $\pm 810$ &      &    &     \\ 
  8  & 172 & $<1.0$ (0.6, 1.4)   & 5740 $\pm 700$ & 3.2 & 28 & $0.13 \pm 0.01$ \\ 
  9  & 230 & 3.0 (2.5, 3.5)   & 4880 $\pm 630$ & & &\\    
  10 & 242 & 3.0  (2.6, 3.6)  & 3840 $\pm 650$ & 5.4 & 15 & $0.53 \pm 0.04$ \\
  12 & 258 & $<2.0$ (1.5, 2.5)   & 4270 $\pm 740$ &     &    & \\
  13 & 319 & 48 (41, 62)      & 1800 $\pm 850$ & 3.9 & 15 & $1.1 \pm 0.03$  \\
  14 & 345 & 22 (20, 26)      & 2700 $\pm 500$ &     &    & \\
  \enddata
  \tablenotetext{a}{Numbering in \cite{katsuda08}.}
  \tablenotetext{b}{Postshock density.}
  \tablenotetext{c}{From \cite{katsuda08}, scaled to a 5 kpc distance.}
  \tablenotetext{d}{Assumed the same as Region 4.}
  \tablenotetext{e}{Between K08 regions 7 and 8 (see Fig.~\ref{rxregions}).}
  \tablecomments{Values for Regions 4, 5, 8, and 12 are determinations, but upper limits
  for densities at the extreme edge.}
  \label{datatable}
\end{deluxetable*}

\begin{deluxetable*}{lcccccccc}
  \tablecolumns{9}
  \tablecaption{Derived Properties of Regions\label{derived}}
  \tablehead{
    \colhead {Region} & $B$ & $u_e$ & $u_B$ & $u_e/u_B$ & $P$ & $\epsilon_e$ 
    & $\epsilon_B$ & $u_e/u_{Bav}$\tablenotemark{a} \\ 
    & ($\mu$G) & ($10^{-10}$ erg cm$^{-3}$) & ($10^{-10}$ erg cm$^{-3}$)
     & & ($10^{-7}$ dyn cm$^{-2}$) & ($10^{-4}$) & ($10^{-4}$) & } 
  \startdata
  3 & 105 & 53    & 4.36  & 12.2   & 7.9     & 68       & 56      &  7.1   \\ 
  4 & 157 & 0.74  & 9.84  & 0.076  & $<11$   & $>0.69$  & $>9.1$  &  0.20  \\ 
  5 & 161 & 2.44  & 10.3  & 0.24   & $<12$   & $>2.0$   & $>8.3$  &  0.68  \\ 
  7 & 121 & 15.9  & 5.8   & 2.75   & 3.1     & 52       & 19      &  2.7   \\ 
  8 & 148 & 2.54  & 8.7   & 0.29   & $<2.1$  & $>12$    & $>42$   &  0.61  \\ 
 10 &  80 & 55.6  & 2.5   & 22     & 2.6     & 220      & 9.9     &  4.7   \\ 
 13 &  60 & 189   & 1.4   & 133    & 9.0     & 210      & 1.6     &  9.7   \\ 
 \enddata
 \tablenotetext{a}{Assuming the median value of $B_{av} = 121\ \mu$G, and
     $u_{Bav} = 5.83 \times 10^{-10}$ erg cm$^{-1}$, for all regions.} 
\end{deluxetable*}

\subsection{Radio intensities}

We determine mean intensities over each small region assuming it
is homogeneous, and that the line-of-sight depth $L$ is equal to the
transverse extent of the filament being sampled (so generally larger than
the region's radial width).  Flux densities in each region are
measured from the 5 GHz radio image \citep{delaney02}, then corrected
for background taken from comparable or larger regions outside the
remnant, divided by the solid angle $\Delta \Omega$, and extrapolated
to 1 GHz assuming $j_\nu \propto \nu^{-0.71}$ everywhere.  The results
are listed as $\langle I_9\rangle$ in Table~\ref{obs}, and the derived
electron energy densities in Table~\ref{derived}.  Quoted
uncertainties result from the off-source rms fluctuation level of 0.14
mJy \citep{delaney02}, scaled by the square root of the extraction
area.

\section{Results}
\label{results}


The globally averaged values for efficiencies for our six remnants are
given in Table~\ref{global}.  The range is extraordinary: a factor of
500 in $\epsilon_e$ and 90 for $\epsilon_B$.  Their ratio (also equal
to $u_e/u_B$), free of uncertainties in $n_0$ and less dependent on
$v_{\rm sh}$, ranges over a factor of 270.

However, the situation worsens when we consider spatial variations in
Kepler's remnant alone.  It can easily be seen without detailed
analysis that the strong north-south brightness gradient in Kepler's
radio emission requires strong spatial variations in one or both
$\epsilon$ factors.  Now $j_\nu \propto K B^{(s+1)/2}$ (Eq.~\ref{jnu})
and $K \propto u_e E_l^{s-2} = \epsilon_e P_2 E_l^{s-2}$
(Eq.~\ref{ue0}), while $B \propto (\epsilon_B P_2)^{1/2}$.  This gives
\begin{equation}
  j_\nu \propto P_2^{(s+5)/4} \epsilon_e \epsilon_B^{(s + 1)/4} E_l^{-(2-s)}.
\end{equation}
The interior of Kepler is in rough pressure equilibrium; that is,
  the post-shock pressure $P_2$ is constant within a factor of a few.
  Certainly $E_l$ is constant.  Then we have no way to explain the
factor of 20 variation in mean intensity recorded in Table~\ref{obs}
other than strong variations of the epsilons.  Certainly the
line-of-sight depth $L$ required to obtain $j_\nu$ from $\langle I_9
\rangle$ is unlikely to vary by this much.

\begin{figure}
  \centerline{\includegraphics[width=3.7truein]{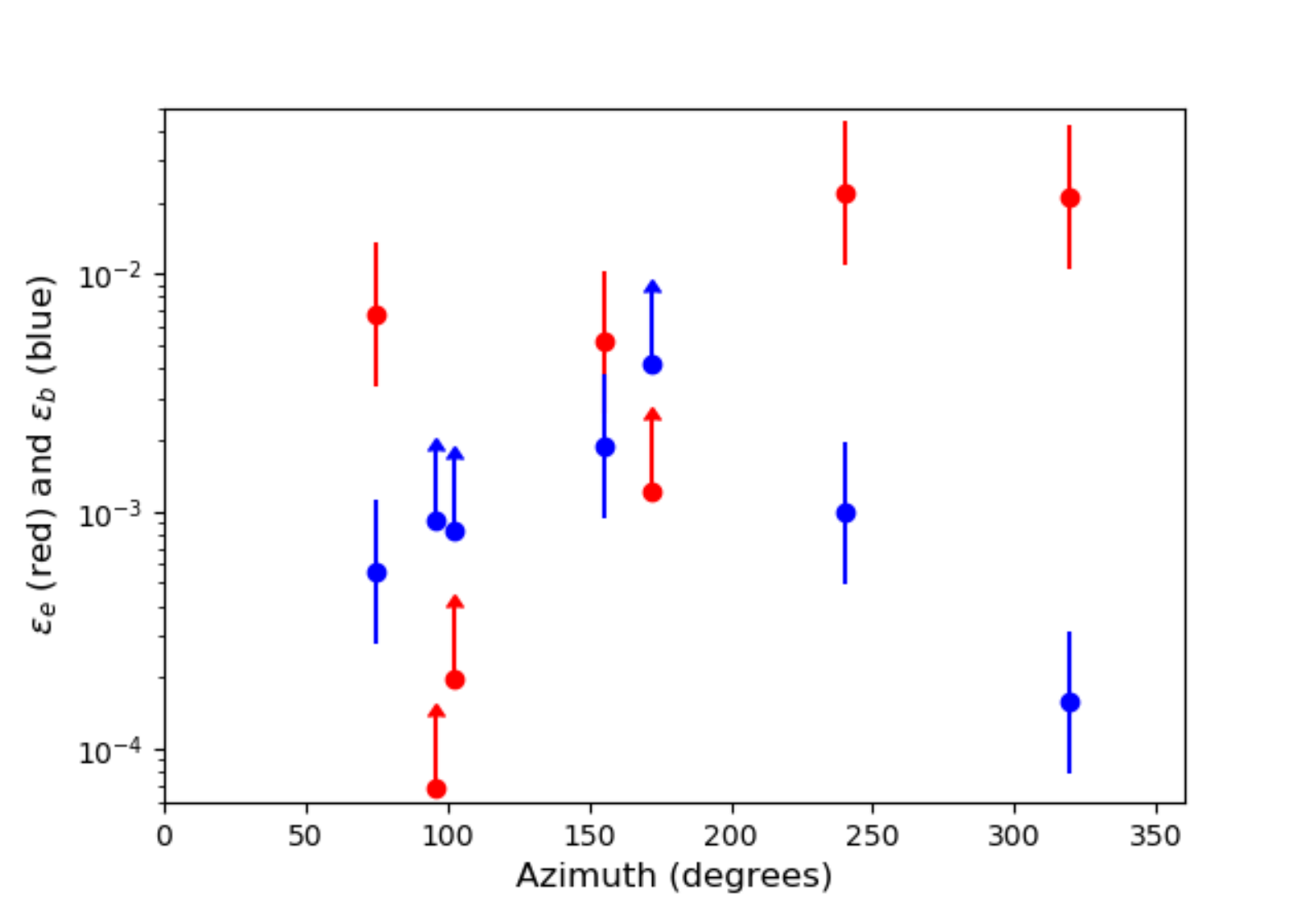}\hskip0.1truein
  \includegraphics[width=3.7truein]{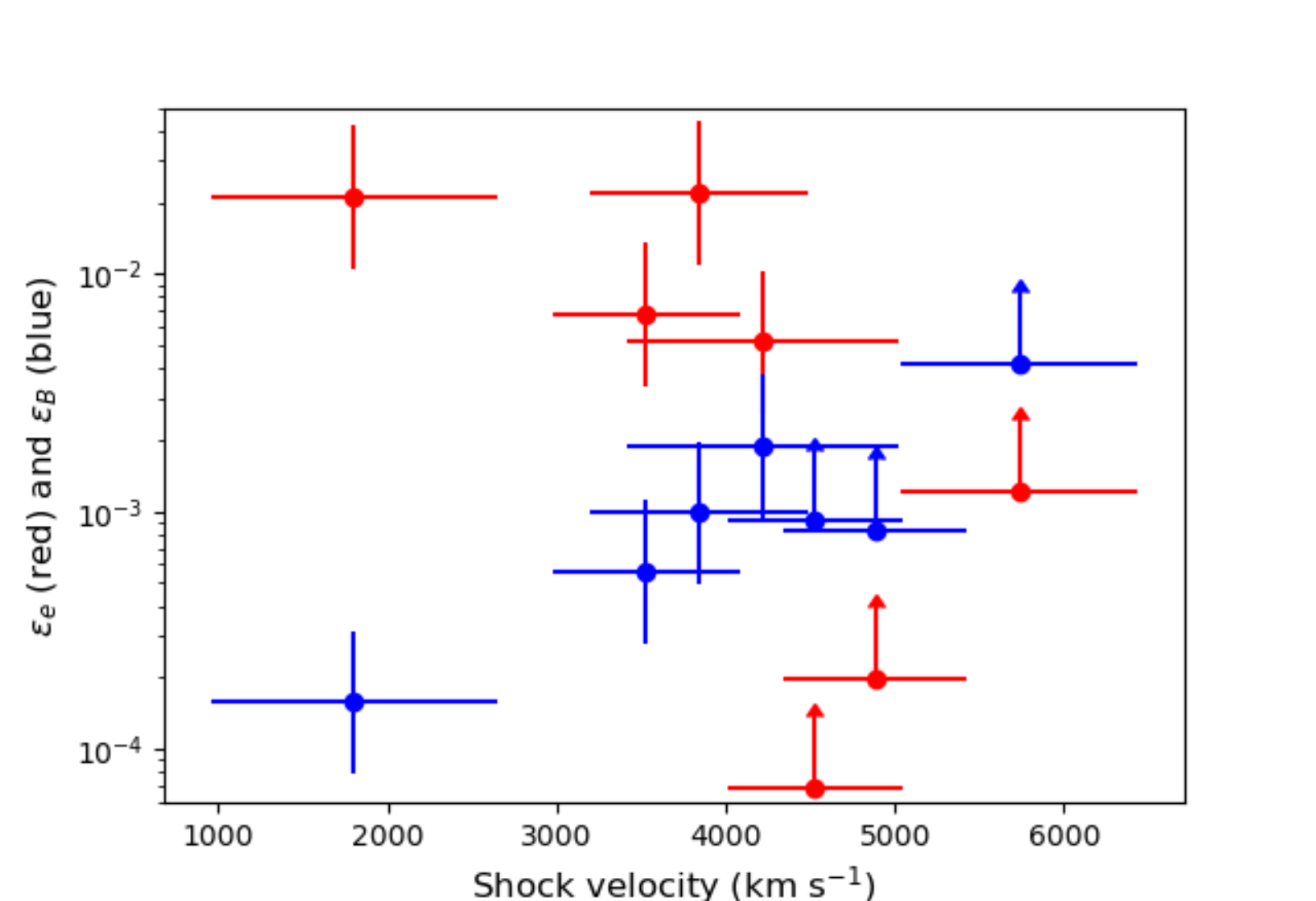}}
  \caption{Left: Efficiencies of relativistic-electron acceleration
    $\epsilon_e$ and magnetic-field amplification $\epsilon_B$ for
    seven locations around the perimeter of Kepler.  The values, and
    their ratio, scatter over orders of magnitude. Right: Efficiences
    vs.~shock velocity.  In both cases, error bars are not formal uncertainties,
  but simply illustrate a range of a factor of 2.  See text. }  
  \label{effics}
  \end{figure}

The quantitative results of Table~\ref{derived}, plotted in
Fig.~\ref{effics}, make this point clearly.  Error bars shown
  there are not formal uncertainties, but simply illustrate an assumed
  factor of 2 (100\%) range.  Section 7.1 discusses uncertainties in
  detail.  But Fig.~\ref{effics} shows that the efficiencies of
magnetic-field amplification and relativistic-electron acceleration
vary by a far greater amount, orders of magnitude, around the
periphery of Kepler.  Their ratio, completely independent of the
pressure determination (and therefore of the density upper limits in
some regions), varies by over three orders of magnitude, both larger
and smaller than 1.  Even removing the two extreme regions 4 and 13,
the range is a factor of 90.  In particular, adjoining regions show no
particular correlation in efficiencies; the smooth trends of density
or pressure with azimuth shown in Fig.~\ref{nh_vs-pr} are not evident.
Figure~\ref{effics} also plots the efficiencies vs.~shock velocity;
while a weak trend appears to be present, it is almost entirely due to
one region, 13, with a much lower shock velocity.

\begin{figure}
  \centerline{\includegraphics[width=3.7truein]{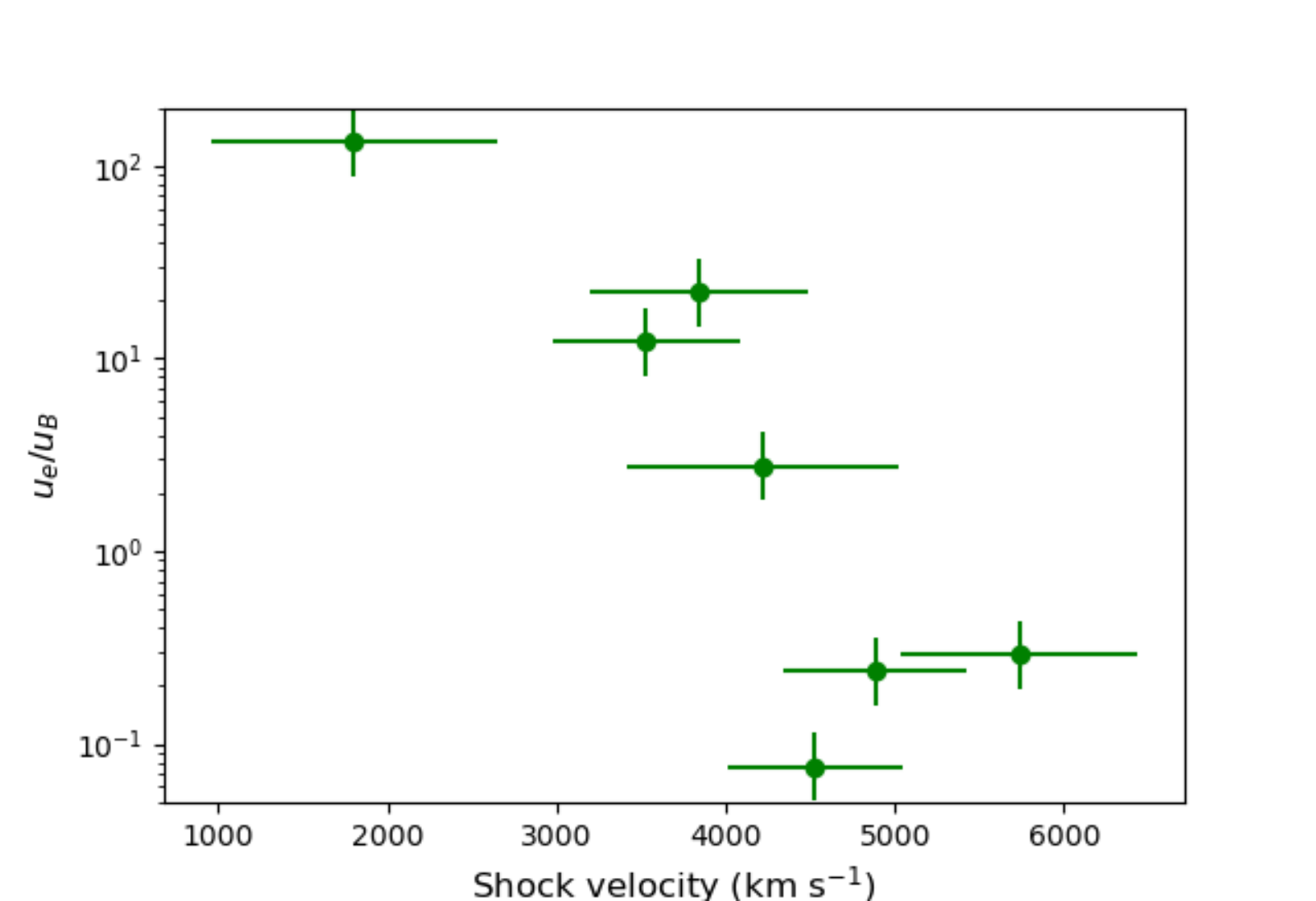}\hskip0.1truein
    \includegraphics[width=3.7truein]{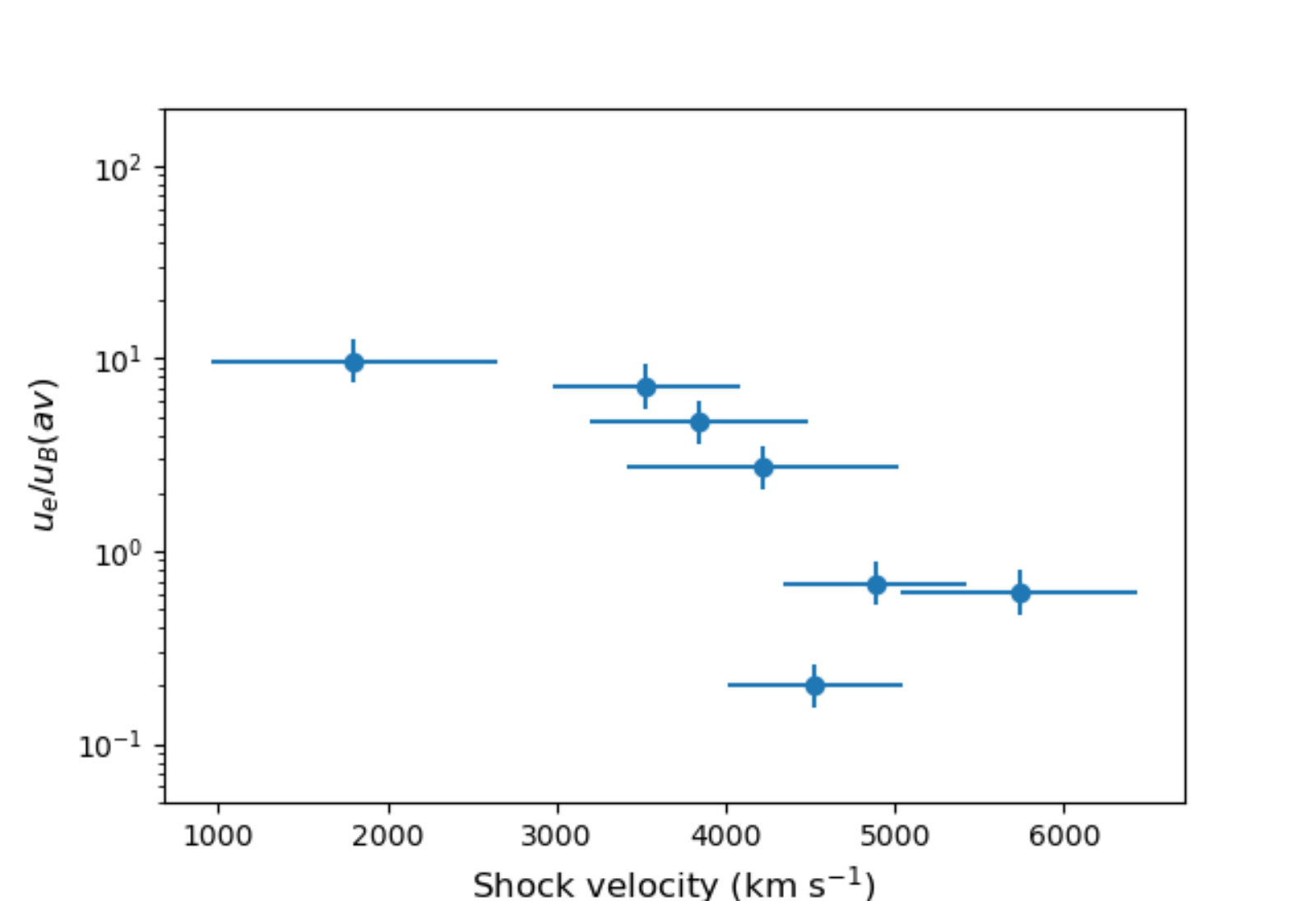}}
  \caption{Left: $u_e/u_B$ vs.~shock velocity.  Right: $u_e/u_{Bav}$
    vs.~shock velocity (i.e., assuming the median-$B$ value to
    calculate both $u_e$ and $u_B$).  Again, error bars are not formal uncertainties,
  but illustrate a range of 50\% in the left panel and 30\% in the right. }
  \label{ratios}
\end{figure}

We also plot the dimensionless ratio $u_e/u_B$ in
  Figure~\ref{ratios}.  These values also scatter widely, ranging over
  three orders of magnitude around Kepler's periphery, with values
  both larger and smaller than 1.
Now the inference of $u_e$ from observations depends strongly on the
inferred magnetic-field strength (Equation~\ref{ue2}), inducing a
strong indirect dependence on shock speed through
Equation~\ref{rimfield}: $u_e \propto B^{-1.71}$.  Any intrinsic
dependence of the observed synchrotron intensity $I_\nu$ on shock
speed can be isolated by plotting the dimensionless ratio
$u_e/u_{Bav}$, that is, using the median value of $B$ to calculate
both $u_e$ and $u_B$.  This quantity, Column 9 of Table~\ref{derived},
is plotted vs.~shock speed in Figure~\ref{ratios}.  As for
$\epsilon_B$, there is the suggestion of a trend with velocity, here
to lower $u_e$ with higher shock velocity; however, the scatter is
large. Figure~\ref{ratios} also illustrates ranges of 50\% for
$u_e/u_B$ (left panel) and 30\$ for $u_e/u_{Bav}$ (right panel).
Since the normalization by post-shock pressure is absent, several of
the uncertainties documented in Section 7.1 are not relevant for
either ratio, while all variation of magnetic field is removed in the
right panel.  Future observational and theoretical studies should
  address the possibility of trends with shock velocity.  Other young
  SNRs such as Tycho are susceptible to a spatially resolved analysis
  such as this.

\section{Discussion}
\label{disc}

Our startling results require closer examination.  First, the values
of $\epsilon_B$, including the value for average remnant properties,
spread over two orders of magnitude for remnant averages, and for
Kepler in particular, are far lower than the 0.1 -- 0.01 often assumed for
nonrelativistic shocks \citep[e.g.,][]{lundqvist20}. A more
sophisticated model for $\epsilon_B$ \citep{sarbadhicary17} predicts
values which, while not constant, range only over a factor of 3.

Most significant for constraining these efficiencies is the range of
values around the rim of Kepler.  Table~\ref{derived} and
Figure~\ref{effics} show a particularly large range (factor of up to 300) in
$\epsilon_e$, while $\epsilon_B$ varies by a smaller, but still large
factor of 35.

\subsection{Possible sources of uncertainty}

The magnitude of the variations among regions shown in
Table~\ref{derived} means that uncertainties of even factors of
several cannot change the qualitative result of strong variations of
efficiency.  However, it is worth considering various possible sources
of uncertainty.

\subsubsection{Distance}

If shock velocities are obtained from angular proper motions,
Equation~\ref{rimfield} shows that magnetic fields inferred from the
angular widths of azimuthal thin rims are independent of distance.
Electron energy densities inferred from Equations~\ref{ue1} and~\ref{ue2}  depend only
on the distance-independent intensity $I_\nu$ and the line-of-sight
depth $L$; if the latter is estimated, as we do here, by the angular
dimensions of the emitting region in the plane of the sky, we have
$u_e \propto d^{-1}$.  Alternatively since $u_e \propto j_\nu$, if
an observed radio flux and angular size are used in Equation~\ref{Snu},
we have $V_{\rm em} \propto d^3$ so again $u_e \propto d^{-1}$.
Thus $u_e/u_B \propto d^{-1}$.  Our method of obtaining the
density at different locations around the rim of Kepler is independent
of distance, so the postshock pressure $P_2 \propto d^2$, giving
$\epsilon_e \propto d^{-3}$ and $\epsilon_B \propto d^{-2}$.

The range of values of $u_e$ in Table~\ref{global} is $\sim 10^4$,
although if Cas A is removed, the range drops to 180.  These values
are so large compared to the factor $\sim 2$ which could be due to
distance uncertainties, that it is hard to imagine that such
uncertainties could mask or remove the dramatic trends of
Table~\ref{global}.  The spread in values of $\epsilon_e$ and
$\epsilon_B$ is considerably smaller, but $\epsilon_e$ also varies by
a factor of 500 (or 51 without Cas A).

\subsubsection{Density determinations}

The quoted uncertainties in densities determined from fitted dust
temperatures reflect primarily the range in assumed proton
temperatures; the statistical errors in the fits are much smaller, for
a particular model of grain composition, structure, and size
distribution.  The models (described in detail in Williams et
al.~2011a) do make assumptions about these quantities, but they are
constrained to some extent by the spectra.  The most generous
allowance for uncertainties in these quantities is unlikely to exceed
a factor of 2 in any case. 

For the regions for which we feel the true immediate post-shock
density may be less than that for the emission we detect and fit in
those regions (regions 4, 5, 8, and 12), there is a possibility that
$\epsilon_e$ is considerably larger than the lower limits shown in
Fig.~\ref{effics}.  Since those are the lowest values, it is
conceivable that the true densities could be low enough to reduce the
scatter considerably.  This would require, however, more than an order
of magnitude difference, reducing the inferred pressure in those
regions to values considerably below those in neighboring regions,
well outside the scatter in pressure found elsewhere, unless
  blast-wave velocities are substantially lower as well.  Such density
  variations could reflect the propagation of the blast wave into an
  inhomogeneous upstream medium, seen in projection. We believe it
unlikely that large variations are present, but without more sensitive
infrared observations at higher spatial resolution, we cannot rule out
the possibility.  In any case, the large scatter in
$\epsilon_e/\epsilon_B$ is, of course, unaffected.

\subsubsection{Shock compression ratio}

The compression ratio enters into the determination of the magnetic
field (Equation~\ref{rimfield}) and into the upstream density as
determined from collisionally heated dust downstream: $B \propto
r^{-2/3}$ and $\rho_0 \propto r^{-1}$.  In numerical results above, we
have assumed a compression ratio of 4.  We know from the presence of
synchrotron rims that the blast wave in Kepler is accelerating
electrons to TeV energies, and it is conceivable that the effective
compression ratio is larger than 4, due either to particle escape or
to an energetically significant population of relativistic ions
\citep{jones91}, but it is very unlikely to be larger than the
relativistic limit of 7, since ample post-shock thermal emission
indicates that the shock is not dominated by highly relativistic
particles.

So $u_B \propto r^{-4/3}$ and $\epsilon_B \propto r^{-1/3}$.  Then
$u_e \propto B^{1 + \alpha} \propto r^{-2(1 + \alpha)/3}$ and $\epsilon_e
\propto r^{(1 - 2\alpha)/3}$.  For our assumed radio spectral index
  $\alpha = 0.71$, we have $u_e \propto r^{-1.14}$ and $\epsilon_e
  \propto r^{-0.14}$.  For $r = 7$, then, we would reduce $u_B$ by
  a factor of 0.47 and $\epsilon_B$ by 0.83, while $u_e$ would drop
  by 0.53 and $\epsilon_e$ by 0.92.  So at most a factor of 2
  uncertainty results from this range of compression ratios.

\subsubsection{Magnetic fields}

As pointed out above, large uncertainties accompany the estimates of
magnetic field from filament widths.  However, all estimates agree on
requiring substantial magnetic-field enhancement over a factor of at
most 4 (or 7) increase due to shock compression alone.  To estimate
the effects of this uncertainty, we have assigned all regions a single
value $B_{\rm av} = 122\ \mu$G, the median of the values shown in
Table~\ref{derived}, which gives $u_B = 5.92 \times 10^{-10}$ erg
cm$^{-3}$.  The last columns of Table~\ref{derived} give $u_e/u_{Bav}$
and $\epsilon_{Bav}$ for this value of $B$. The spread in $u_e/u_B$
is reduced from about 1800 to 50, which is the spread in $u_e$ alone.

\subsubsection{Radio properties}

The radio spectral index of Kepler between 1.4 and 5 GHz is observed
to vary with location between about $0.65$ and $0.8$ over most of the
remnant, including all of the regions we have measured
\citep{delaney02}.  This corresponds to a range in electron energy
index $s$ of 2.3 to 2.6.  In terms of $\alpha = (s-1)/2$, the
spectral-index dependencies of the electron energy density $u_e$ are
\begin{equation}
  u_e \propto E_l^{1-2\alpha}(2\alpha - 1)^{-1} c_j(\alpha) B^{-(1 + \alpha)}\nu^\alpha.
  \end{equation}
Evaluating this expression for the median magnetic field of 122 $\mu$G, $E_l = 8.2 \times 10^{-7}$
erg, and $\nu = 1$ GHz, for $\alpha = 0.65$ and 0.8, gives
\begin{equation}
  \frac{u_e(\alpha = 0.8)}{u_e(\alpha = 0.65)} = 2.4.
\end{equation}
Very little of the scatter in the values of $u_e$ at different locations in Kepler can be
due to variations of $\alpha.$

It is also unlikely that huge variations in the minimum energy of the
electron distribution occur at different locations.  Our fiducial
value of $10 m_e c^2$ simply reflects the energy at which electrons
are relativistic enough for the synchrotron formulae on which our
analysis rests to apply.

The radio flux measurements for the different regions carry
uncertainties, as listed in Table~\ref{obs}.  Those were obtained
from the off-source rms values of 4.8 GHz radio flux in the image of
\cite{delaney02}, scaled by the square root of the extraction area.
The S/N ratio for the regions ranges from 3 for region 3 to 31 for
region 13.  These cannot explain the orders of magnitude spread
evident in Table~\ref{derived} and exhibited in Figure~\ref{effics}.

Finally, we obtain emissivities by assuming a line-of-sight depth of
our regions, which we take to be the longer dimension of our extraction
region.  Again, this can easily be in error by a factor of a few, but
not by orders of magnitude.

Unfortunately, we cannot calculate formal uncertainties in $u_e$ or
$u_B$, or their ratio.  While we report uncertainties in density and
shock velocity, these cannot be propagated to functions of those
quantities without knowledge of their distributions, quite unlikely to
be Gaussian.  We know even less about the uncertainties in magnetic
field.  The variations among authors illustrated in Table 1 are not
statistical or systematic errors, but results of applications of
slightly different versions of the basic argument embodied in
Equation~\ref{rimfield}.  A better estimate is represented by the
spread shown in Table 1 for the reference whose values we have used,
\cite{parizot06} -- of order 10\%.  But again, we do not know enough
about the distribution of the uncertainties to be able to apply a
simple error propagation formalism.  Figures 6 and 7 illustrate what
different levels of uncertainty would look like in the distributions;
the values we have chosen represent our estimates of reasonable ranges
based on the discussion above.

\subsection{Consequences}

We are forced to conclude that the various sources of uncertainty are
dwarfed by the enormous spread of values of $u_e$, $u_B$,
$\epsilon_e$, and $\epsilon_B$ at different locations on Kepler's
periphery.  The primary source of the spread in $u_e/u_B$ results from
the radio brightness variations (Equation~\ref{ue2}).  We conjecture
that the missing physics required to explain the spread has to do with
acceleration of particles, electrons in particular, which evidently
has a more complex dependence on parameters than we have included. No
monotonic relation between $u_e$ and shock velocity is apparent in the
values in Tables~\ref{obs} and~\ref{derived}; another likely
possibility, the shock obliquity angle $\theta_{\rm Bn}$ between the
shock velocity and upstream magnetic field, may have a highly
nonlinear effect on the ultimate population of $\sim$ GeV electrons
producing the radio emission.

What has become of equipartition, the time-honored principle still
often used to infer properties of synchrotron sources?  First, in the
SNR context, SNRs are very inefficient at producing either magnetic
field energy or relativistic-particle energy.  Compared, say, to
extragalactic radio sources, relatively little of the total SN energy
ever winds up in nonthermal forms (that is, both epsilons are always
small).  It is easier to imagine wide variations in $u_e/u_B$ when a
much larger pool of thermal energy is available for any of various
processes to produce electrons or magnetic field.  In addition,
absence of direct information on relativistic protons means that we
simply have no idea what the total nonthermal energy density is
(though very broad inferences from observed Galactic cosmic rays seem
to require about 10\% of total SN energy winding up in relativistic
baryons).  So electron energy densities are a small fraction of a
fairly small fraction, and we should not be astonished if there is no
clear relation between the energy in relativistic electrons and other
pools of energy. As for magnetic energy, when SNR magnetic fields are
inferred using equipartition arguments, they essentially serve as
proxies for the SNR mean surface brightness $\Sigma$, as can be seen
from, for instance, Pacholczyk (1970), Equation 7.14, where $B_{\rm
  equip} \propto (\Sigma/D)^{2/7}$, $D$ being the source diameter.
There is no obvious mechanism which could operate to transfer energy
to or from magnetic field based on the energy in relativistic
electrons.  Fig.~\ref{ratios} shows that the density-independent
  ratio $u_e/u_B$ varies by orders of magnitude in regions of comparable
  shock velocity, emphasizing this point.

A related consequence of our results for Kepler is the problematic
nature of the traditional $\Sigma-D$ relation as a diagnostic tool.
The large azimuthal variations we find suggest that global averages
can be very misleading.  While the average values of $\epsilon_e$ and
$\epsilon_B$ we obtain from the global properties for Kepler
(Table~\ref{global}) do fall within the large range of values at
particular locations, it is not clear that those global values
represent any kind of mean of the physical properties.  Great care
should be taken in drawing conclusions based on such global values.

One might hope to extract from this exercise some clue as to what
additional parameters might be required to explain the range of
efficiencies.  The absence of smooth trends of efficiencies with position
angle around the remnant disfavors explanations such as a smoothly
varying shock obliquity angle $\theta_{\rm Bn}$, as one might
expect for a blast wave encountering a uniform upstream magnetic
field.  The large scatter for adjoining regions shown in
Fig.~\ref{effics} requires relatively small-scale variations in
properties important for particle acceleration and/or magnetic-field
amplification.  Our attempt to extract information relatively
independent of magnetic-field inferences is shown in
Fig.~\ref{ratios}, right panel, where a noisy trend is visible as
lower electron-energy density with higher shock velocity.  We do not
claim the unambiguous existence of such a relation since the apparent
trend is strongly dependent on one or two points, but the data might
offer a clue toward identifying the additional physics evidently
necessary to fully understand electron acceleration in strong shock
waves.

\section{Conclusions}
\label{concls}

We have used a combination of radio, infrared, and X-ray data to
estimate nonthermal energy densities in relativistic electrons and
magnetic field first, in global averages over six young supernova
remnants, and second, for seven regions around the periphery of
Kepler's supernova remnant.  For the various remnants, we find
enormous ranges of these quantities, with $u_e/u_B$ varying by a
factor of over 3000 (or among the Type Ia remnants, i.e., excluding
Cas A, by a factor of 70).  The efficiencies range from $10^{-4}$ to
0.05 for $\epsilon_e$ and from 0.002 to 0.1 for $u_B$.  There are no
clear trends with age or shock speed evident in Table~\ref{global}.

Of course, the collection of remnants is quite inhomogeneous, and the
objects themselves are inhomogeneous, so describing them by global
average values may conceal systematic trends in these quantities.  We
have therefore extracted densities, shock speeds, magnetic-field
strengths, and radio fluxes from seven regions around Kepler, and find
similarly large spreads in the energy densities, their ratio, and the
efficiencies.  Figures~\ref{effics} and~\ref{ratios} summarize the
results.

The variations in our inferred values are so large that even
conservative assessments of sources of error or uncertainty are quite
unable to account for them.  There seems no alternative but to
conclude that additional parameters must inflluence the amplification
of magnetic field and acceleration of electrons.  Various
possibilities come to mind: the neutral fraction of upstream gas, the
obliquity angle between the local shock normal and the upstream
magnetic-field direction, or downstream variations in, for instance,
magnetic turbulence.  Evidence for local variations of
shock-acceleration physics at different points around SNR peripheries
(through the dimensionless diffusion coefficient $\eta \equiv
\lambda_{\rm mfp}/r_L$, the ratio of mean free path to Larmor radius)
has recently been presented by \cite{tsuji21}; the origins of the
variations of $\eta$ they find presumably rely on some of these
additional parameters.

Given our evident lack of understanding of all the factors that
contribute to shock acceleration of particles and turbulent
amplification of magnetic field, it would seem that assumptions of,
for instance, constant values of the efficiency as a function of time
or local conditions in synchrotron sources should be made with extreme
caution, and borne in mind as possible sources of systematic error or
completely incorrect inference when analyzing spectra of such objects
as gamma-ray bursters or radio supernovae.

\acknowledgments This work was supported by NASA grants NNX11AB14G
through the ADAP program and award RSA 1378040 through the {\sl
  Spitzer} program.  We thank Tracey De Laney for making available
the Kepler radio data.  A preliminary version of this work was presented
in \cite{reynolds12b}.

\bibliography{snrk}
\end{document}